\documentclass[a4paper,fleqn]{cas-sc}

\usepackage[authoryear,longnamesfirst]{natbib}
\usepackage{subcaption}
\usepackage{graphicx}
\usepackage{subcaption}
 \usepackage{setspace}
\usepackage{textcomp}     
\usepackage{url}
\usepackage[]{amsmath}
\usepackage{amsmath}
\usepackage{nccmath}
\usepackage{lineno}
\usepackage{longtable}
\usepackage{float,lscape}
\usepackage{layout}
\usepackage[flushleft]{threeparttable}
\usepackage{siunitx}

\newcommand{\an}{{\textquotesingle Ayl\'{o}\textquotesingle chaxnim }}
\newcommand{\anns}{{\textquotesingle Ayl\'{o}\textquotesingle chaxnim}}

\begin{document}
\let\WriteBookmarks\relax
\def\floatpagepagefraction{1}
\def\textpagefraction{.001}

\shorttitle{Palomar twilight survey of \anns, Atiras, and comets}

\shortauthors{Bolin et al.}  

  \title[mode = title]{The Palomar twilight survey of \anns, Atiras, and comets}
\author[]{B. T. Bolin$^{a,b,c,*,**}$}[orcid=0000-0002-4950-6323]
\author[3]{F. J. Masci}[]
\author[4]{M. W. Coughlin}[]
\author[5]{D. A. Duev}[]
\author[19]{\v{Z}. Ivezi\'{c}}[]
\author[21]{R. L. Jones}[]
\author[19]{P. Yoachim}[]
\author[2,6]{T. Ahumada}[]
\author[7]{V. Bhalerao}[]
\author[7]{H. Choudhary}[]
\author[8]{C. Contreras}[]
\author[9,10]{Y.-C. Cheng}[]
\author[11]{C. M. Copperwheat}[]
\author[12,13]{K. Deshmukh}[]
\author[2,14]{C. Fremling}[]
\author[15,16]{M. Granvik}[]
\author[17]{K. K. Hardegree-Ullman}[]
\author[18]{A. Y. Q. Ho}[]
\author[20]{R. Jedicke}[]
\author[2]{M. Kasliwal}[]
\author[7]{H. Kumar}[]
\author[22]{Z.-Y. Lin}[]
\author[2]{A. Mahabal}[]
\author[17]{A. Monson}[]
\author[2,14]{J. D. Neill}[]
\author[23]{D. Nesvorn\'{y}}[]
\author[11]{D. A. Perley}[]
\author[14]{J. N. Purdum}[]
\author[24,25]{R. Quimby}[]
\author[26]{E. Serabyn}[]
\author[2,7]{K. Sharma}[]
\author[7]{V. Swain}[]
\address[1]{Goddard Space Flight Center, 8800 Greenbelt Road, Greenbelt, MD 20771, USA}
\address[2]{Division of Physics, Mathematics and Astronomy, California Institute of Technology, Pasadena, CA 91125, USA}
\address[3]{IPAC, California Institute of Technology, Pasadena, CA 91125, USA}
\address[4]{School of Physics and Astronomy, University of Minnesota, Twin Cities, Minneapolis, MN 55455, USA}
\address[5]{Weights $\&$ Biases, San Francisco, CA 94103, USA}
\address[6]{Department of Astronomy, University of Maryland, College Park, MD 20740, USA}
\address[7]{ Department of Physics, Indian Institute of Technology Bombay, Powai, 400 076, India}
\address[8]{Las Campanas Observatory, Casilla 601, La Serena, Chile}
\address[9]{Department of Physics, National Taiwan Normal University, Taipei City 116325, Taiwan}
\address[10]{Center of Astronomy and Gravitation, National Taiwan Normal University, Taipei City 116325, Taiwan}
\address[11]{Astrophysics Research Institute, IC2, Liverpool Science Park, 146 Brownlow Hill, Liverpool L3 5RF, UK}
\address[12]{Institute of Astronomy, Katholieke Universiteit Leuven, 3000 Leuven, Belgium}
\address[13]{Department of Metallurgical Engineering and Materials Science, Indian Institute of Technology Bombay, Powai, 400 076, India}
\address[14]{Caltech Optical Observatories, California Institute of Technology, Pasadena, CA 91125, USA}
\address[15]{Department of Physics, University of Helsinki, Helsinki 00014, Finland}
\address[16]{Asteroid Engineering Lab, Lule\r{a} University of Technology,  Kiruna SE-981 28, Sweden}
\address[17]{Steward Observatory, University of Arizona, Tucson, AZ 85721, USA}
\address[18]{Department of Astronomy, Cornell University, Ithaca, NY 14853, USA}
\address[19]{Department of Astronomy and DiRAC Institute, University of Washington, Seattle, WA 98195, USA}
\address[20]{Institute for Astronomy, University of Hawai'i, 2680 Woodlawn Dr., Honolulu, HI 96822, USA}
\address[21]{Aerotek and Rubin Observatory, Tucson, AZ, USA}
\address[22]{Department of Physics, National Taiwan Normal University, Taipei City 116325, Taiwan}
\address[23]{Department of Space Studies, Southwest Research Institute, Boulder, CO 80302, USA}
\address[24]{Department of Astronomy, San Diego State University, 5500 Campanile Dr, San Diego, CA 92182, U.S.A.}
\address[25]{Kavli Institute for the Physics and Mathematics of the Universe (WPI), The University of Tokyo Institutes for Advanced Study, The University of Tokyo, Kashiwa, Chiba 277-8583, Japan}
\address[26]{Jet Propulsion Laboratory, California Institute of Technology, Pasadena, CA 91109, USA}
\cortext[cor1]{Corresponding author: bolin.astro@gmail.com}
\cortext[cor2]{NASA Postdoctoral Program Fellow.} 

\begin{abstract}[S U M M A R Y]
Near-sun sky twilight observations allow for the detection of asteroids interior to the orbit of Venus (Aylos) and the Earth (Atiras) and comets. We present the results of observations with the Palomar 48-inch telescope (P48)/Zwicky Transient Facility (ZTF) camera in 30 s r-band exposures taken during evening astronomical twilight from 2019 Sep 20 to 2022 March 7 and during morning astronomical twilight sky from 2019 Sep 21 to 2022 Sep 29. More than 21,940 exposures were taken in evening astronomical twilight within 31$^{\circ}$ and 66$^{\circ}$ from the Sun with an r-band limiting magnitude between 18.0 and 20.8 (5th to 95th percentile), and more than 24,370 exposures were taken in morning astronomical twilight within 31$^{\circ}$  and 65$^{\circ}$ from the Sun with an r-band limiting magnitude between 18.2 and 20.9 (5th to 95th percentile). The morning and evening twilight pointings show a slight seasonal dependence in limiting magnitude and ability to point closer towards the Sun, with limiting magnitude improving by ~0.5 magnitudes during the summer months and Sun-centric angular distances as small as ~31-32$^{\circ}$ during the spring and fall months. In total, the one Aylo, (594913) \anns, and 4 Atiras, 2020 OV$_1$, 2021 BS$_1$, 2021 PB$_2$, and 2021 VR$_3$, were discovered in evening and morning twilight observations. Additional twilight survey discoveries also include 6 long period comets: C/2020 T2, C/2020 V2, C/2021 D2, C/2021 E3, C/2022 E3 and C/2022 P3, and two short period comets: P/2021 N1 and P/2022 P2 using deep learning comet detection pipelines. The P48/ZTF twilight survey also recovered 11 known Atiras, one Aylo, three short period comes, two long period comets, one interstellar object, 45,536 Main Belt asteroids, and 265 near-Earth objects. Additionally, observations from the GROWTH network of telescopes were used to recover the Aylo, Atira, and comet discoveries made during the ZTF twilight survey. Lastly, we discuss the future twilight surveys for the discovery of Aylos such as with the Vera Rubin Observatory which will have a twilight survey starting in its first year of operations and will cover the sky as within 45 degrees from the Sun. Twilight surveys such as those by ZTF and future surveys will provide opportunities for the discovery of asteroids inside the orbits of the terrestrial planets that would otherwise be unavailable in conventional sky survey observations.
\end{abstract}

\maketitle
\section{Introduction}
Near-Earth asteroids are the diffused members of the Main Belt asteroid population that have escaped through various resonances found throughout the Main Belt \citep[][]{Binzel2015,Granvik2017}. A subset of near-Earth asteroids (NEAs), Atira asteroids are found on orbits with aphelia, 0.718 au $<$ $Q$ $<$ 0.983 au, just outside the aphelion of Venus and inside the perihelion distance of the Earth \citep[][]{Granvik2018,Nesvorny2023NEOMOD}. \an asteroids have aphelia $Q$ $<$ 0.718 au, inhabiting the orbital space inside the perihelion of Venus. Hereafter, we use the abbreviation `Aylo' for the \an class of asteroids. Models describing the population of NEAs predict that there are $\sim$10,000 Atira asteroids and $\sim$2,000 Aylo asteroids with absolute magnitude 17 $<$ $H$ $<$ 25 ($\sim$1500 m to $\sim$40 m assuming an albedo of 0.12). \citep[][]{Nesvorny2023NEOMOD}. Searches for asteroids inside the orbits of the Earth and Venus may provide tests of asteroid populations models. For example, the discovery of the first Aylo, \anns \citep[][]{Bolin2020MPECb,Bolin2022IVO}, may suggest that the observed population of Aylo objects in the inner-Solar system may be larger than predicted by contemporary asteroid population models \citep[][]{Bolin2022IVO,Bolin2023Com}. However, a thorough analysis of the Aylo completeness and survey coverage is beyond the scope of this work and reserved for future study.

The observed orbital and physical properties of NEAs are strongly affected by observational selection effects such as limiting magnitude, sky coverage, and cadence \citep[][]{Jedicke2016}. Differences in the properties of different NEA surveys can result in some surveys being advantageous over others towards finding certain types of NEAs, i.e., surveys with fainter limiting magnitudes finding more small objects, etc. \citep[][]{Jedicke2015}. The observed orbital properties of NEAs also have a directional dependence, where the detection of asteroids becomes more probable at low ecliptic latitudes due to the low inclinations of many Main Belt asteroids between 2.0 and 3.5 au from the Sun \citep[][]{Gladman2009}. The detection of asteroids on higher inclinations may require observation of parts of the sky significantly higher above the ecliptic, as for Kuiper Belt objects $>$30 au from the Sun with inclinations $>$10$^{\circ}$ \citep[][]{Petit2017}. NEAs may have a broad ecliptic latitude distribution due to their broader inclination distribution and $<$0.3 au proximity with the Earth \citep[][]{Jedicke2002}. NEAs surveys may cover sky close to the ecliptic plane, resulting in more distant objects being detected as well as detecting objects passing closer to or in orbit around the Earth in searches covering latitudes off of the ecliptic \citep[][]{Granvik2016,Bolin2020CD3,Nesvorny2023NEOMOD}.

The angular distance between NEA survey pointings and the Sun also affects the orbital range of asteroids observed in the survey. For asteroids interior to the orbit of the Earth, the maximum angular distance between the Sun and an asteroid is given by arcsin$\mathrm{\left(Q/1 au\right)}$ where Q is the aphelion of the asteroid. As regions of the sky with ecliptic longitudes closer to the Sun are observed, the minimum Q of asteroids located in the sky that is detectable by the survey decreases \citep[][]{Jedicke2016}. For asteroids with aphelia interior to the $\sim$0.98 au perihelion of Earth, this angle is 90$^{\circ}$, inside the $\sim$0.718 au perihelion of Venus, 46$^{\circ}$, inside the $\sim$0.47 au aphelion of Mercury, $\sim$27.8$^{\circ}$ and inside the $\sim$0.31 au perihelion of Mercury, $\sim$17.9$^{\circ}$. For ground-based surveys, the small angular distance between asteroids interior to the orbit of the Earth and the Sun necessitates observations taking place within $\sim$1 h of sunset or sunrise when sky within this angular distance from the Sun is accessible within the elevation range of most survey telescopes \citep[][]{Jedicke2015}.

Concepts for ground-based surveys designed for the detection of NEAs interior to the orbit of the Earth were demonstrated with the University of Hawaii 88-inch telescope (UH 88-inch) \citep[][]{Whiteley1998} resulting in the discovery of the first known interior-Earth asteroid, 1998 DK$_{36}$ \citep[][]{Tholen1998}. Additionally, simulations of ground-based surveys showed that asteroids interior to the orbit of Venus could be observed during twilight \citep[][]{Masi2003}. Searches for asteroids interor to the orbit of the Earth were conducted by the Lincoln Near-Earth Asteroid Research project \citep[LINEAR, which discovered the first interior-Earth asteroid given a number and a name, (163693) Atira,][]{Stokes2000,Green2003}, Catalina Sky Survey \citep[][]{Zavodny2008}, Pan-STARRS \citep[][]{Denneau2013}, the Isaac Newton Telescope \citep[]{Vaduvescu2017,Vaduvescu2018}, the Palomar 48-inch/Zwicky Transient Facility (P48/ZTF) \cite[e.g.,][]{Bolin2022IVO,Bolin2023Com}. These ground based surveys covered the sky near or interior to quadrature allowing for the detection of inner-Earth asteroids \cite{Jedicke2016}. Space-based efforts to detect asteroids interior to the orbit of the Earth have also been made through the Near Earth Object Surveillance Satellite Mission (NEOSSat) \citep[][]{Hildebrand2004}. Prior to September 2019, 19 Atira objects and no Aylo asteroids were known \citep[][]{mpcatira2024}. Recent bespoke searches for asteroids interior to the orbit of the Earth have seen some success, such as with the V\'{i}ctor M. Blanco 4-m Telescope, which has discovered several new Atira asteroids since 2020 \citep[][]{Sheppard2022atira}.

In addition to Aylos and Atiras, the twilight sky is an ideal location to hunt for comets, especially morning twilight, since it is the location of the sky leading the Sun providing fresh sky to detect objects that have not been observed in the previous months due to their conjunction with the Sun \citep[][]{Jedicke2016}. Recent high-profile discoveries of comets made during morning twilight include interstellar comet 2I/Borisov, which was discovered in 2019 September \citep[][]{Williams2019c} and C/2023 P1 (Nishimura), which was discovered in 2023 August \citep[][]{Nishimura2023MPEC}. As a bonus, the discovery of comets during morning twilight also provides additional time for follow up observers to study them since they are leading the Sun and will take longer to go back into solar conjunction compared to if they had been discovered in evening twilight \citep[][]{Bolin20202I,Bolin2024E3}. A preliminary search for NEAs interior to the orbit of the Earth at Palomar occured for several months between 2018 November and 2019 June using a different cadence, a different data analysis pipeline, and did not use a dedicated network of follow up telescopes for the recovery of candidates \citep[][]{Ye2020rr}.

This manuscript summarizes the asteroid and comet results from the twilight observations made with the P48/ZTF that occurred between 2019 September to 2022 September. A small portion of these observations has been described in previous publications for the discovery of \an \citep[][]{Bolin2020MPECb,Bolin2022IVO} and C/2022 E3 (ZTF) \citep[][]{Bolin2024E3} and will be summarized in this manuscript for completeness. In addition to summarizing P48/ZTF twilight observations, we will describe upcoming twilight observation efforts being designed and implemented into the Rubin Observatory Legacy Survey of Space and Time starting within the first year of its operations.

\section{ZTF Twilight Survey and Follow Up}
\subsection{ZTF Twilight Observations}
The P48 telescope with its ZTF possesses a field of view of 47 sq. deg. and a 5-$\sigma$ limiting magnitude of r$\sim$21 \citep[][]{Bellm2019}. The P48 is primarily used to survey the entire observable night sky in two bands, g (mean wavelength $\sim$483.0 nm) and r-band (mean wavelength $\sim$646.8 nm), to detect transients such as supernovae, variable stars, and electromagnetic components of gravitational wave phenomena \citep[][]{Graham2019}. The ZTF camera consists of 16 separate 6144~pixel $\times$ 6160~pixel Charge-Coupled Device (CCD) arrays on a single camera mounted at the prime focus of the P48.  The plate scale of the camera is 1.01 arcseconds pixel$^{-1}$ has a square 7.4$^{\circ}$  $\times$ 7.4$^{\circ}$  dimensions with an 86.7$\%$ fill factor \citep[][]{Dekany2020}. 
 
Image data are processed with an image differencing pipeline using reference frames to remove static sources. Solar system transients are extracted from the images using the ZTF Moving Object Detection Engine (ZMODE) algorithm for objects moving slower than $\sim$2~$\mathrm{\left(^{\circ}/d\right)}$ \citep[][]{Masci2019}. A special version of ZMODE, called ZMODE Lite, was configured to run on the twilight survey day on a per night, per session basis. The ZMODE Lite pipeline operates by linking pairs of detections found on subsequent images moving at similar velocities. ZMODE Lite was tested on background Main Belt asteroids found in Twilight Survey fields and determined to have a detection efficiency has been shown to be $\gtrsim$80$\%$ for asteroids with r$<$19-20, and a 43$\%$ efficiency, half the maximum value of 87$\%$, of r$<$20.5  as seen in Fig.~1F of \citep{Bolin2023Com} assuming a conversion of V-r of 0.2. This is a significant increase in efficiency compared to twilight observations by \citep[][]{Ye2020rr} in which only V$\sim$19.5 or brighter were detected with an efficiency of 36$\%$, half their maximum efficiency of 72$\%$. The increase in efficiency of our survey compared to \citep[][]{Ye2020rr} is likely due to the increase in the sky coverage, and a cadence and moving object detection and data processing techniques more beneficial to the detection of asteroids. Additionally, \citep[][]{Ye2020rr} is quoted as using ZMODE for their identification of moving objects, however, a different moving object detection algorithm was actually used instead of ZMODE (private communication).

Additionally, the Tails deep-learning neural network is used to identify comets on individual twilight survey images \citep[][]{Duev2021}. We used a 32 CPU core virtual machine instance on Google Cloud computing resources which took $\sim$5-10 h to process a night's worth of images. Algorithms for shift-and-stack detection of solar system transients \citep[][]{Shao2014,Whidden2019} as well as citizen science projects for the detection of comets \citep[][]{Chandler2024} exist, but their application to the twilight survey is beyond the scope of this manuscript. 
  
A small portion of the total P48/ZTF time is allocated to microsurveys for the detection of solar system objects \citep[][]{Chang2022,Yeager2022,Bolin2023Com}. One of the microsurveys is designed to look at the sky with solar distance angles $<$60$^{\circ}$ during nautical and astronomical twilight to detect comets and asteroids inside the orbit of Earth as well as Earth co-orbitals \citep[][]{Bolin2022IVO,Yeager2023}. Between September 2019 and 2022 Sep 29, twilight observations consisted of 5-10$\%$ of the total scheduled ZTF survey time took place during 551 evening nautical twilight sessions and 601 morning nautical twilight sessions. The fraction of ZTF for the twilight survey that were successfully taken was $\sim$5$\%$. The evening and morning portions of the twilight survey ran every night or every other night depending on weather and availability of the telescope. This was a significant increase in cadence compared with earlier twilight observations by \cite{Ye2020rr} which mostly ran every third night between November 2018 and June 2019.

The time available for the twilight survey varies due to the changing length of the night as throughout the year. Fig.~1 shows the amount of time in minutes of 5.5$\%$ of the time between the start of evening nautical twilight and the end of morning nautical twilight as a function of the day of the year. The duration of twilight and the amount corresponding to 5.5$\%$ of the total ZTF survey time are anti-correlated, with the duration of twilight lasting $\sim$38 minutes during the Summer months and decreasing to 28 minutes in the winter months while the amount of time available for the twilight survey peaks at $\sim$36 minutes during the winter months and reduces to $\sim$21 minutes during the summer months. The duration of the twilight survey is adjusted to take into account these seasonal variations so that they only consist of the  5-10$\%$ of the total duration of the night through the year rather than the full nautical to astronomical twilight duration.

\begin{figure}\centering
\includegraphics[width=0.5\linewidth]{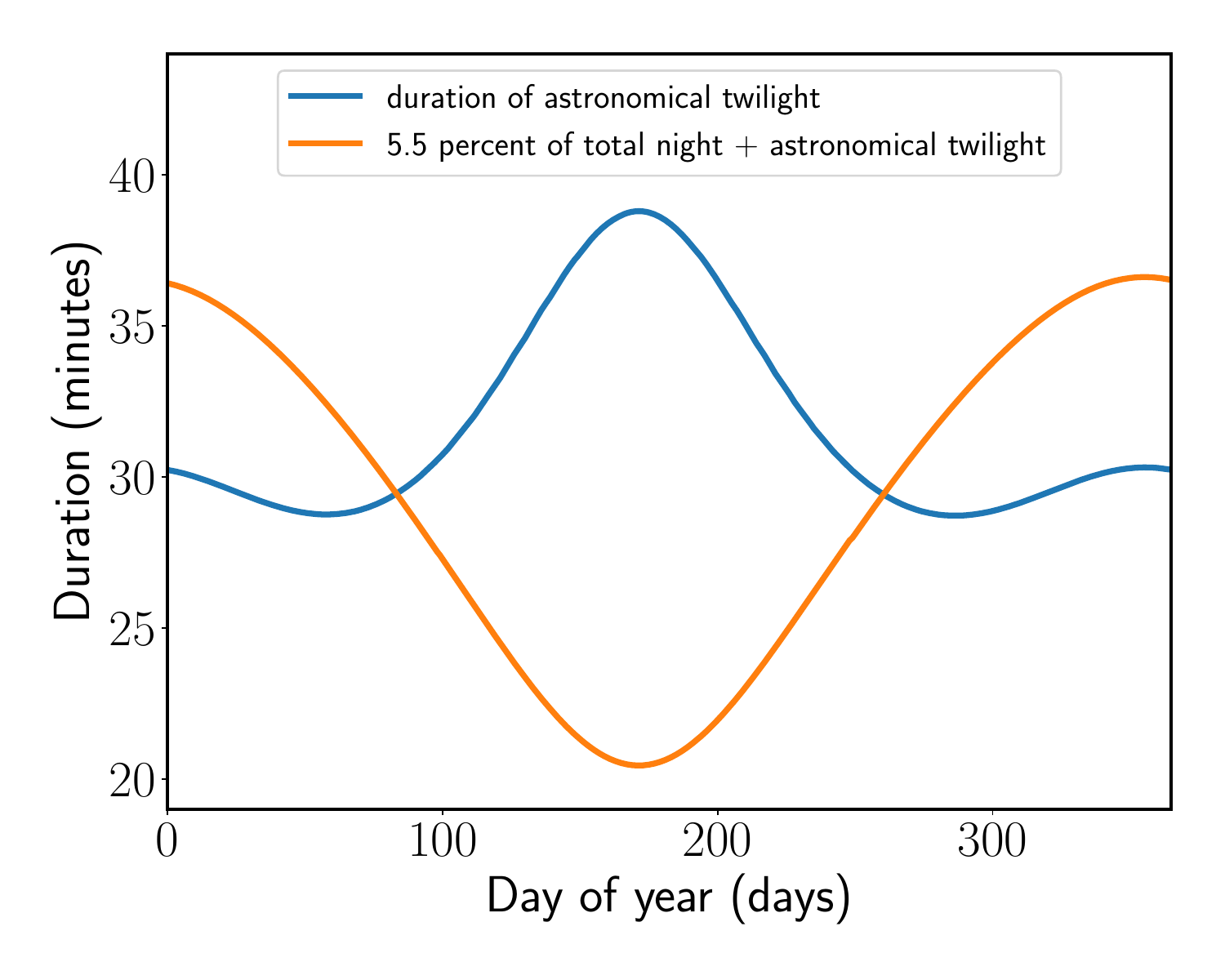} 
\caption{\textbf{Duration of twilight and 5.5$\%$ of twilight + night survey time.} The blue line shows the time between nautical and astronomical twilight (-12$^{\circ}$ to -18$^{\circ}$ Sun altitude) as a function of the day of the year. The orange line shows the time corresponding to 5.5$\%$ of the total survey time including evening/morning nautical twilight and total night time.}
\end{figure}

The Twilight survey uses the ZTF scheduler to arrange the time of the fields limiting the observations to between -12$^{\circ}$ and -18$^{\circ}$ twilight and selecting fields as close to the Sun's direction as possible while keeping the fields above an elevation limit of 20$^{\circ}$ \citep[][]{Bellm2019b, Bolin2022IVO}.  Between 2019 September and 2020 March, the twilight survey occurred on consecutive nights, alternating between evening and morning twilight. On average, each twilight survey session consisted of 10 footprints with 4 x 30 s r-band exposures with an $\sim$10 s readout time for each and a median time between exposures of $\sim$15 s with the telescope slewing to adjacent fields after finishing an exposure. The telescope slewed between exposures with the average time separating individual subsequent exposures on each field was 3-7 minutes. From 2020 March to 2022 September, the time allocated to twilight survey was increased from $\sim$5$\%$ to 10$\%$ when the survey was executed during each night during evening and morning twilight. In addition, the number of exposures per footprint was increased from 4 to 5 reducing the sky coverage but increasing the number of detections per footprint to result in higher accuracy of moving object linkages between subsequent nights \citep[][]{Masci2019}.

The total duration between the first exposure and the last exposure in the 4 exposures per foot print scheme was $\sim$10 to 15 minutes which was extended to 15 to 20 minutes in the 5 exposures per footprint scheme. The top panel of Fig.~2 shows the Sun-centric ecliptic sky-plane distribution for twilight survey fields between 2019 September and 2021 April. The majority of the twilight survey pointings occurred at $>$20$^{\circ}$ Sun-centric ecliptic latitude and $>$$\left | 20 \right |$$^{\circ}$ Sun-centric ecliptic longitude. The order of observations between footprints was optimized to reduce the slew time between fields and the slew time for an entire survey session.

\begin{figure}\centering
\includegraphics[width=0.7\linewidth]{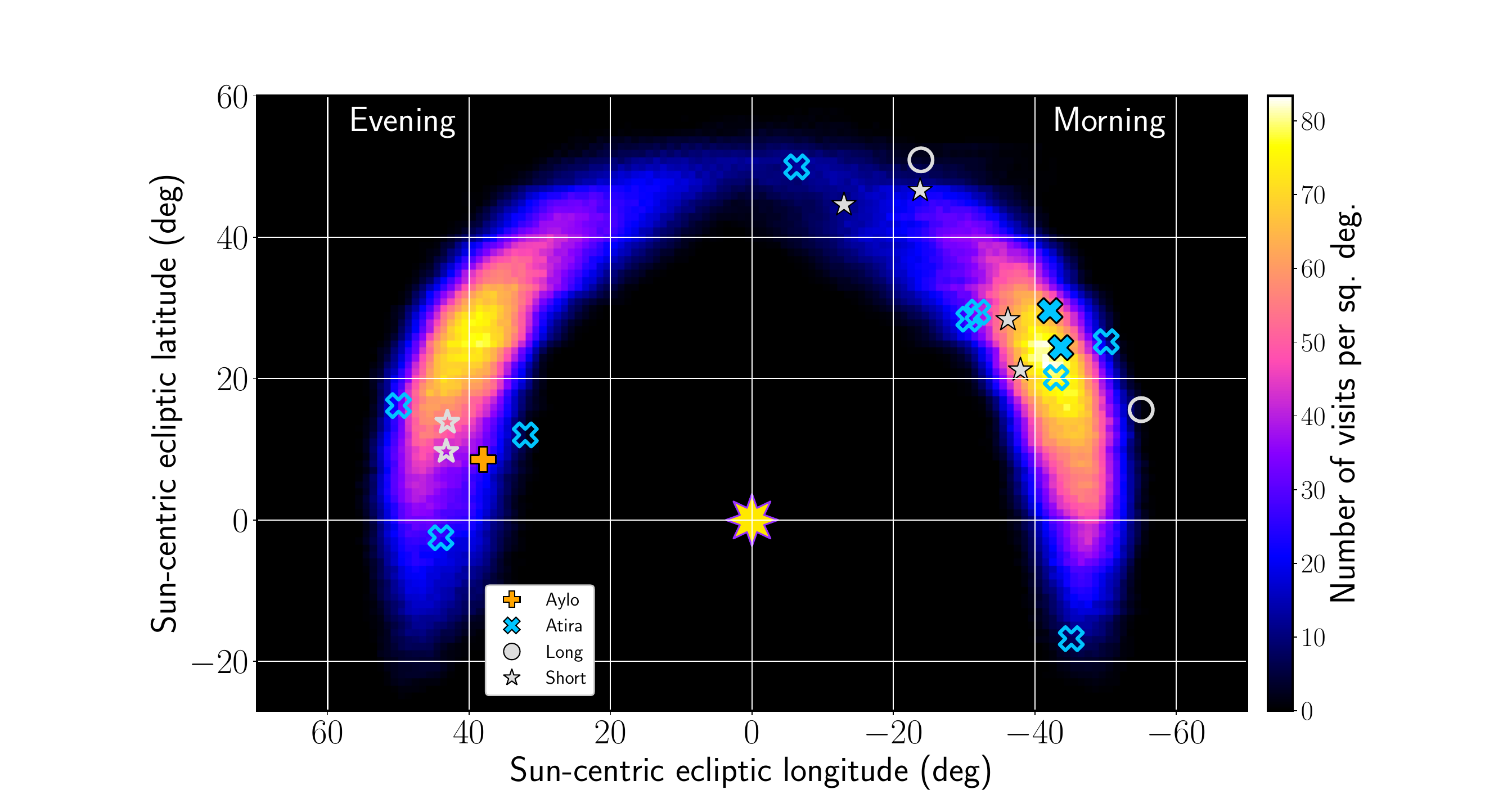} 
\includegraphics[width=0.7\linewidth]{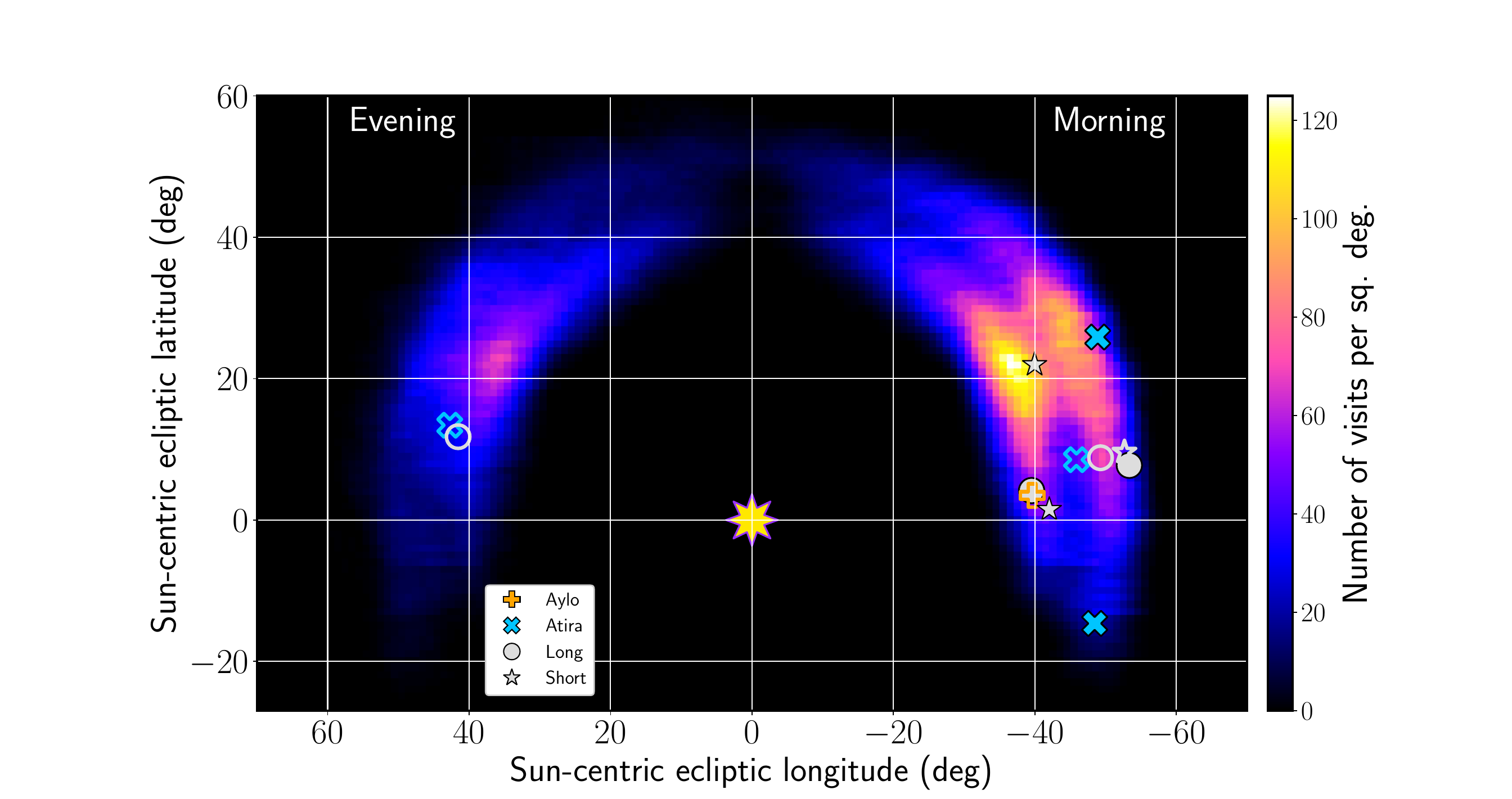}
\includegraphics[width=0.7\linewidth]{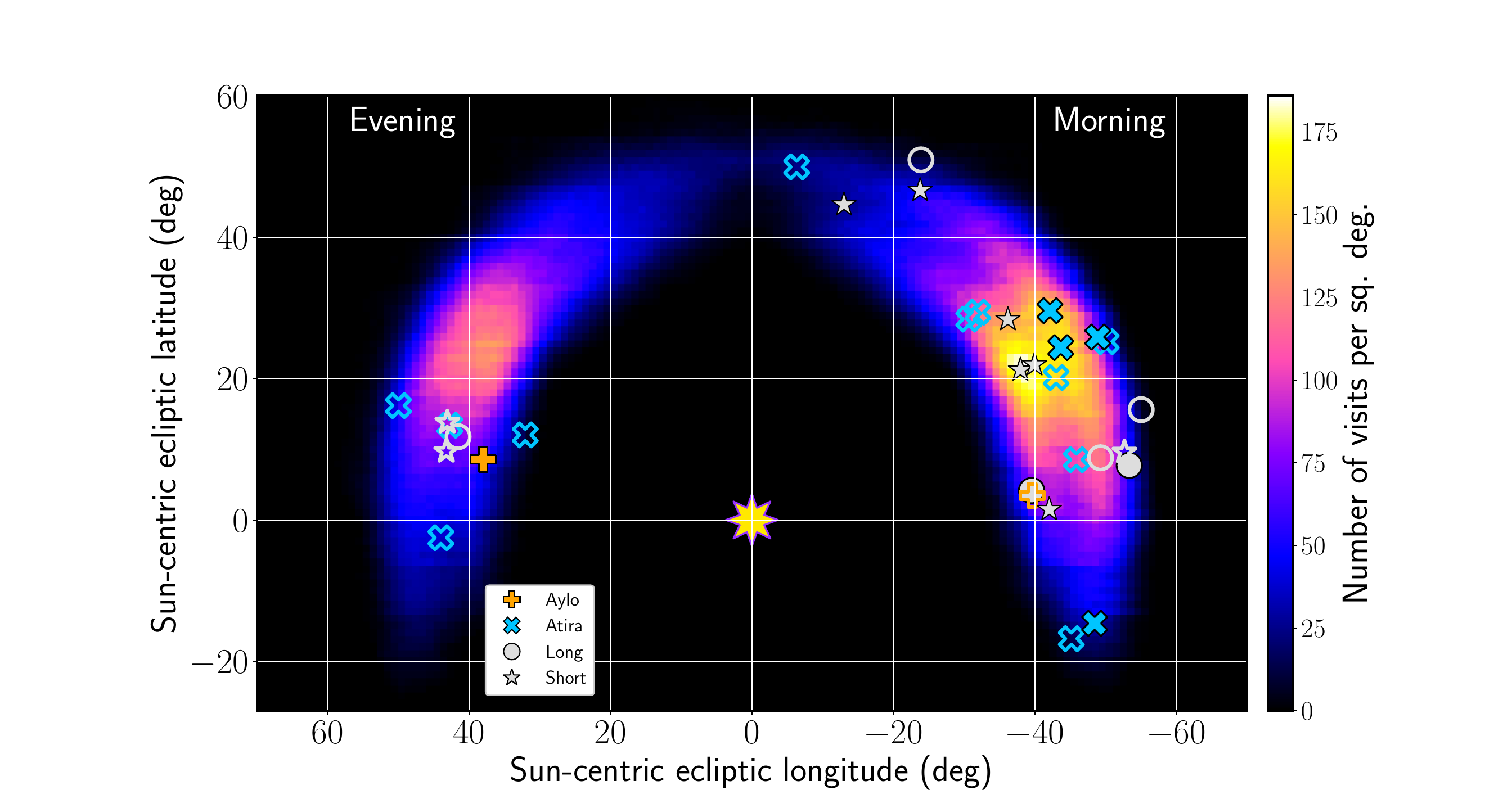} 
\caption{\textbf{Sky coverage of the ZTF twilight survey between 2019 September and 2022 September.} \textbf{Top panel:} the Sun-centric ecliptic sky-plane distribution of ZTF twilight survey fields taken between 2019 September and 2021 April. \textbf{Center panel:} the skyplane distribution of ZTF twilight survey fields taken between 2021 April and 2022 March for the evening portion and 2021 April and 2022 September for the morning portion. \textbf{Bottom panel:} the skyplane distribution of ZTF twilight survey fields taken between 2019 September and 2022 March for the evening portion and 2019 September and 2022 September for the morning portion. The Sun-centric sky plane position of Aylo and Atira asteroids and long and short-period comets are indicated by crosses, x's, circles, and stars. These symbols are filled in for ZTF discoveries and outlined for recoveries. The color bar indicates the number of visits per sq.deg. covered by the twilight survey. The position of the Sun is indicated by a yellow starburst.}
\end{figure}

Starting in 2021 April the total number of footprints in a given twilight survey session was varied to take into account the variable duration of the night throughout the year while keeping the number of exposures per footprint to five. During the winter, a total of 13 footprints were observed during a survey session, while only a total of 7 were observed during the Summer. In addition, the twilight survey fields were divided into a ``low group'' and a ``high group'' to cover parts of the sky closer to the Sun more efficiently and cover the sky closer to the quadrature. The low group consisted of five fields grouped and observed five times each with 30 s r-band exposures in alternating order, leaving 3-5 minutes between exposures on the same footprint.  The Sun-centric angular distance distribution of twilight survey fields after 2021 April is bimodal as seen in the top left and top right panels of Fig.~3. The average Sun-centric angular distance of the low fields was $\sim$40$^{\circ}$. The Sun-centric angular distance is calculated over the entire area covered by the ZTF camera footprint on the sky. The high group consisted of the remainder of the fields after the first five fields were observed and had a a 3-5 minute separation between exposures. The average Sun-centric angular distance of the high group was $\sim$50$^{\circ}$. 

The center panel of Fig.~2 shows the skyplane distribution of the twilight survey fields after the twilight survey fields were organized into the high and low groups in 2021 April. The bifurcation between the high and low fields is evident in both the evening and morning sessions, with the low fields being concentrated at $\sim$20$^{\circ}$ Sun-centric ecliptic latitude and $<$$\left | 40 \right |$$^{\circ}$ Sun-centric ecliptic longitude. The bifurcation results in smaller Sun-centric angular distance coverage as seen in the top right panel of Fig.~3 showing the cumulative Sun-centric angular distance distribution subsequent to the bifurcation compared to the Sun-centric angular distance distributions shown for the complete coverage of the survey between 2019 September to 2022 September, especially for Sun-centric angular distances of less than 40$^{\circ}$ as seen in the center right, bottom left and bottom right panels of Fig.~3. The bottom panel of Fig.~2 shows the combined Sun-centric skyplane distribution of all evening and morning twilight survey fields taken between 2019 September and 2022 March for the evening portion and 2019 September and 2022 September for the morning portion.

\begin{figure}\centering
\includegraphics[width=0.4\linewidth]{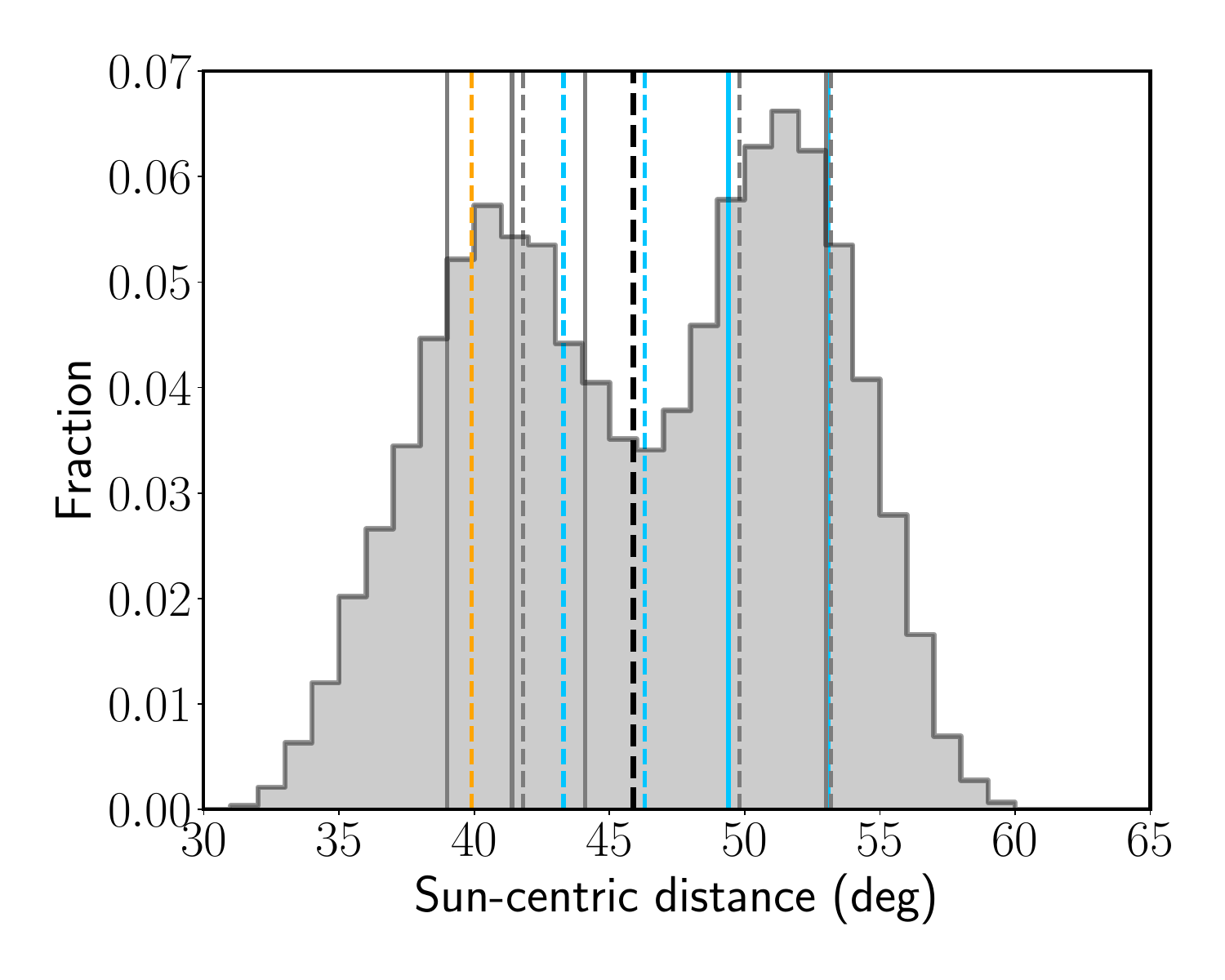}
\includegraphics[width=0.4\linewidth]{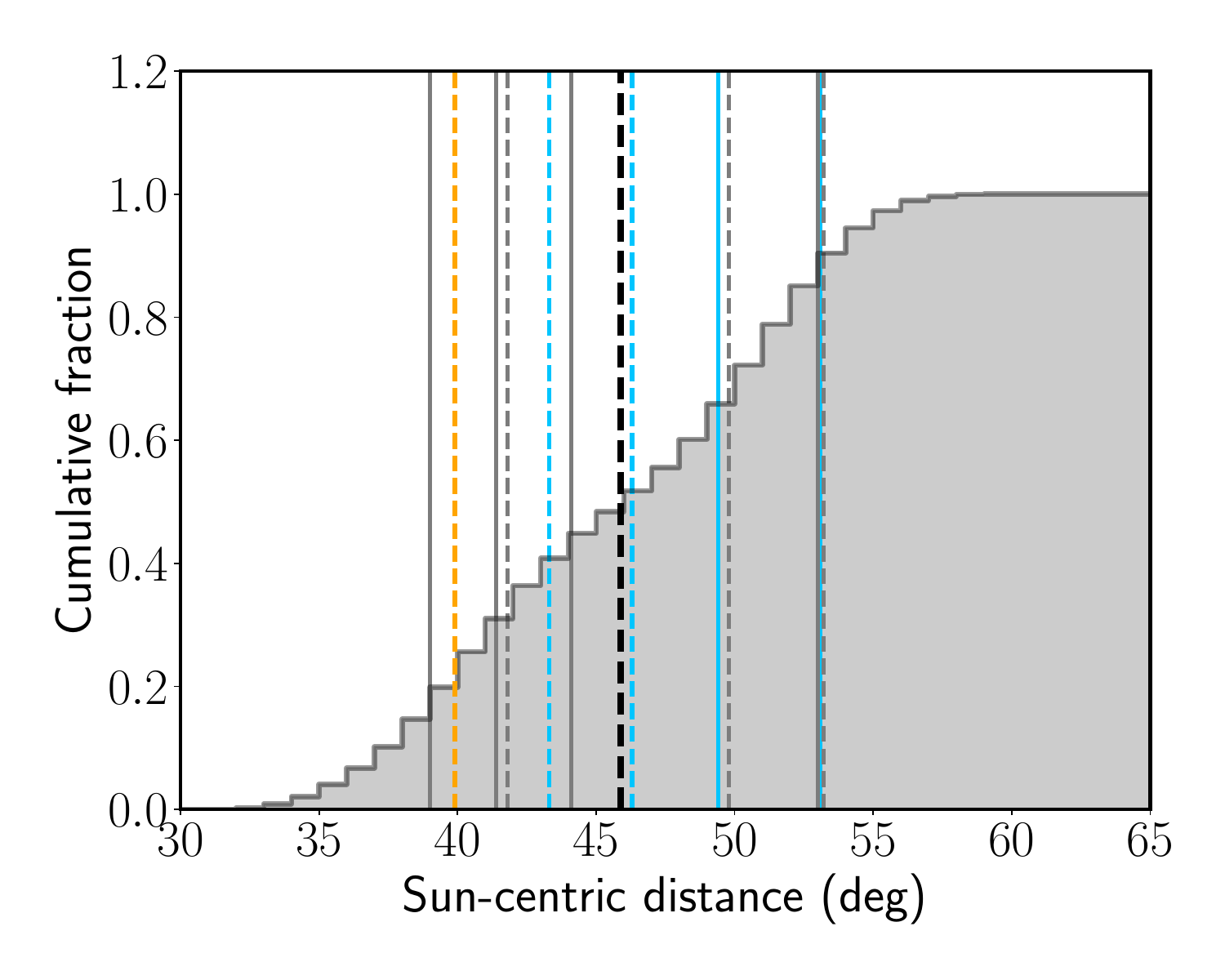} 
\includegraphics[width=0.4\linewidth]{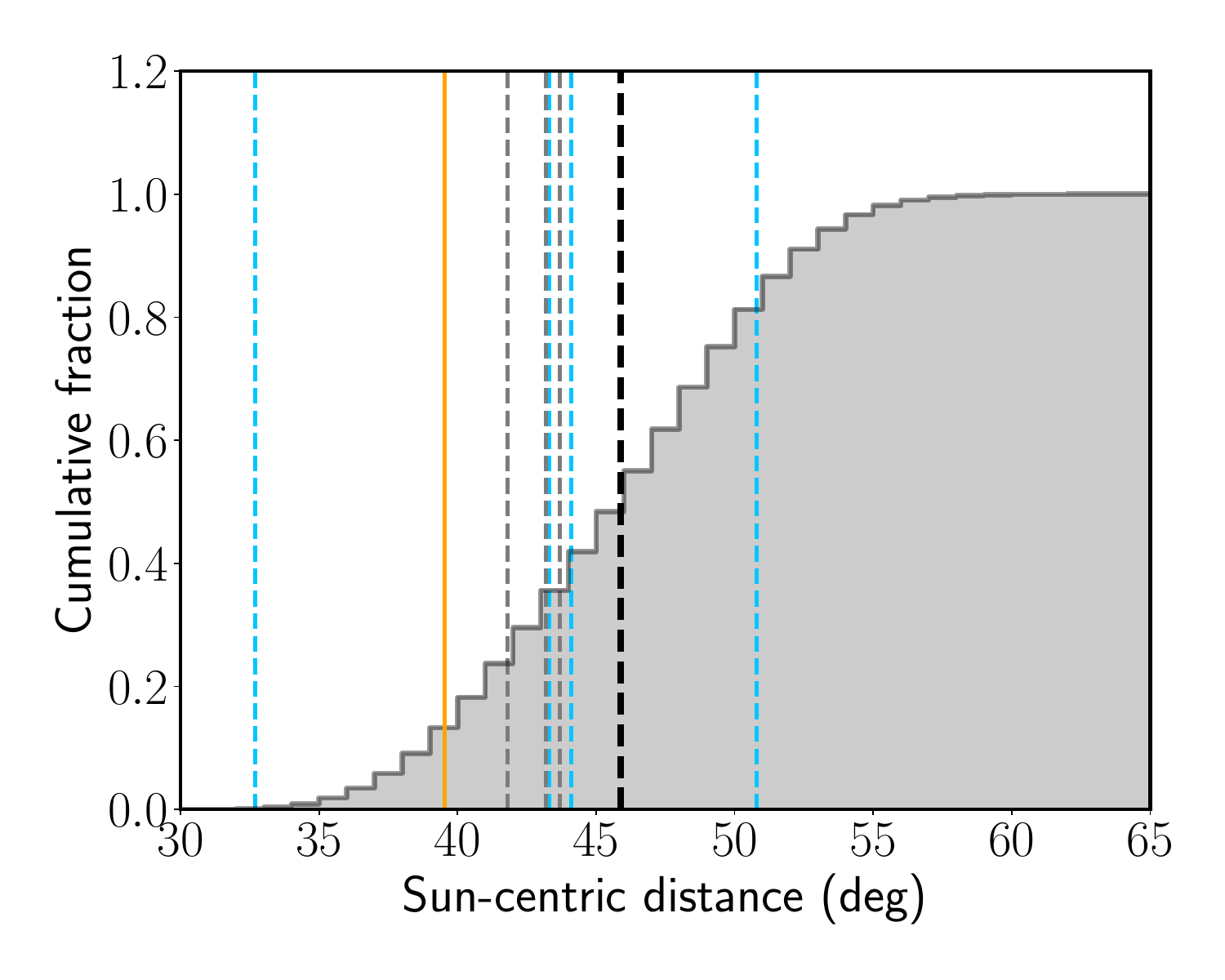} 
\includegraphics[width=0.4\linewidth]{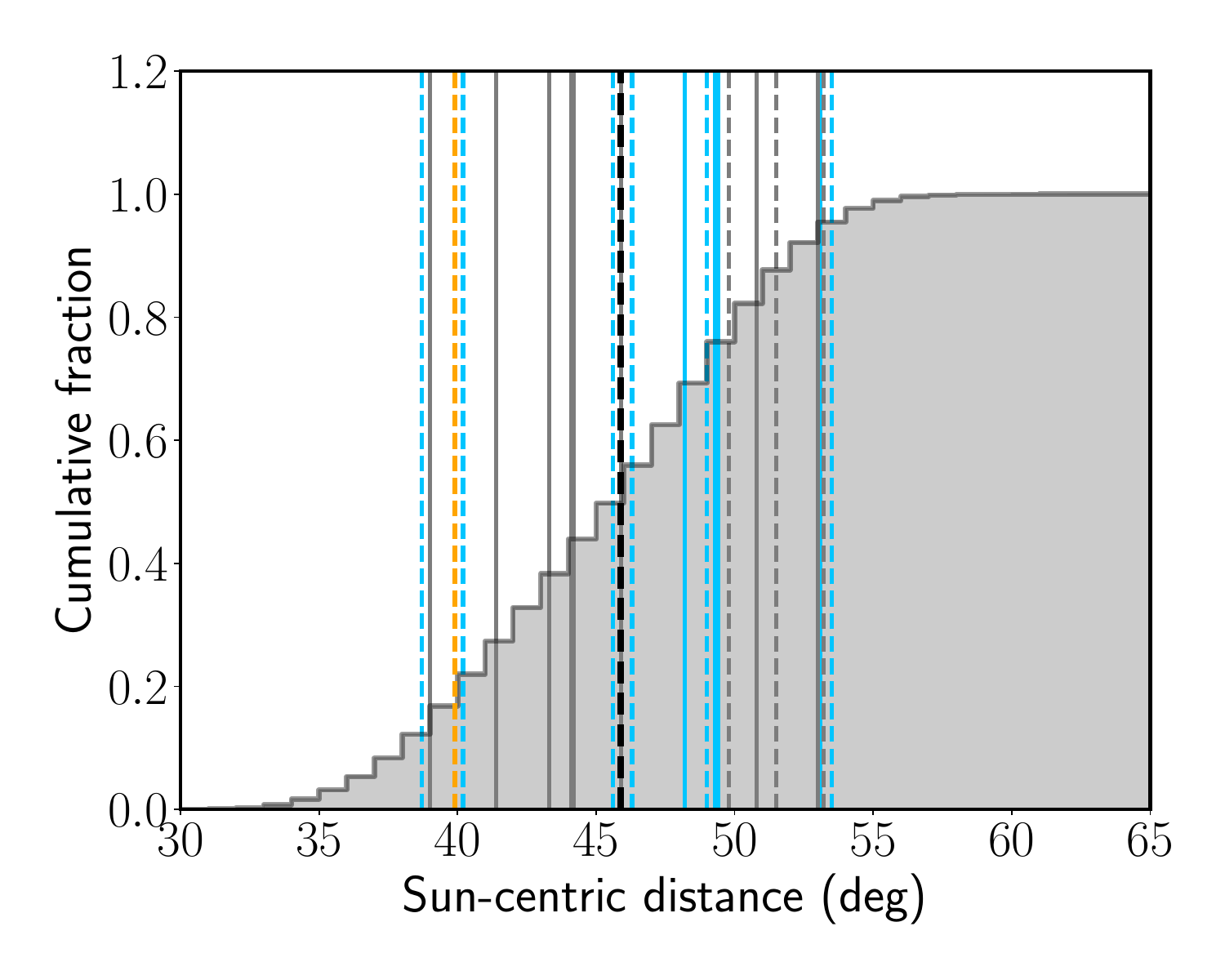}
\includegraphics[width=0.4\linewidth]{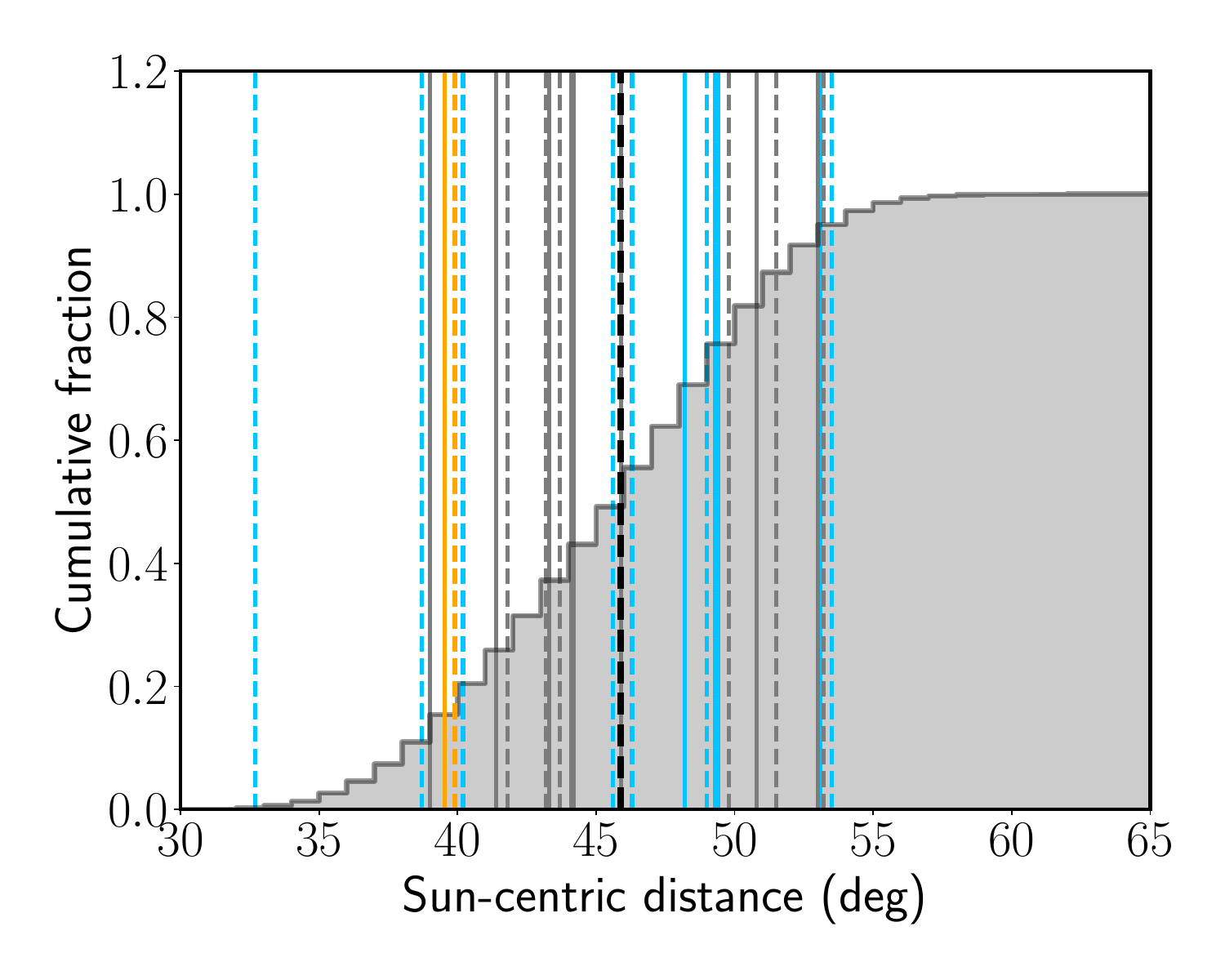} 
\includegraphics[width=0.4\linewidth]{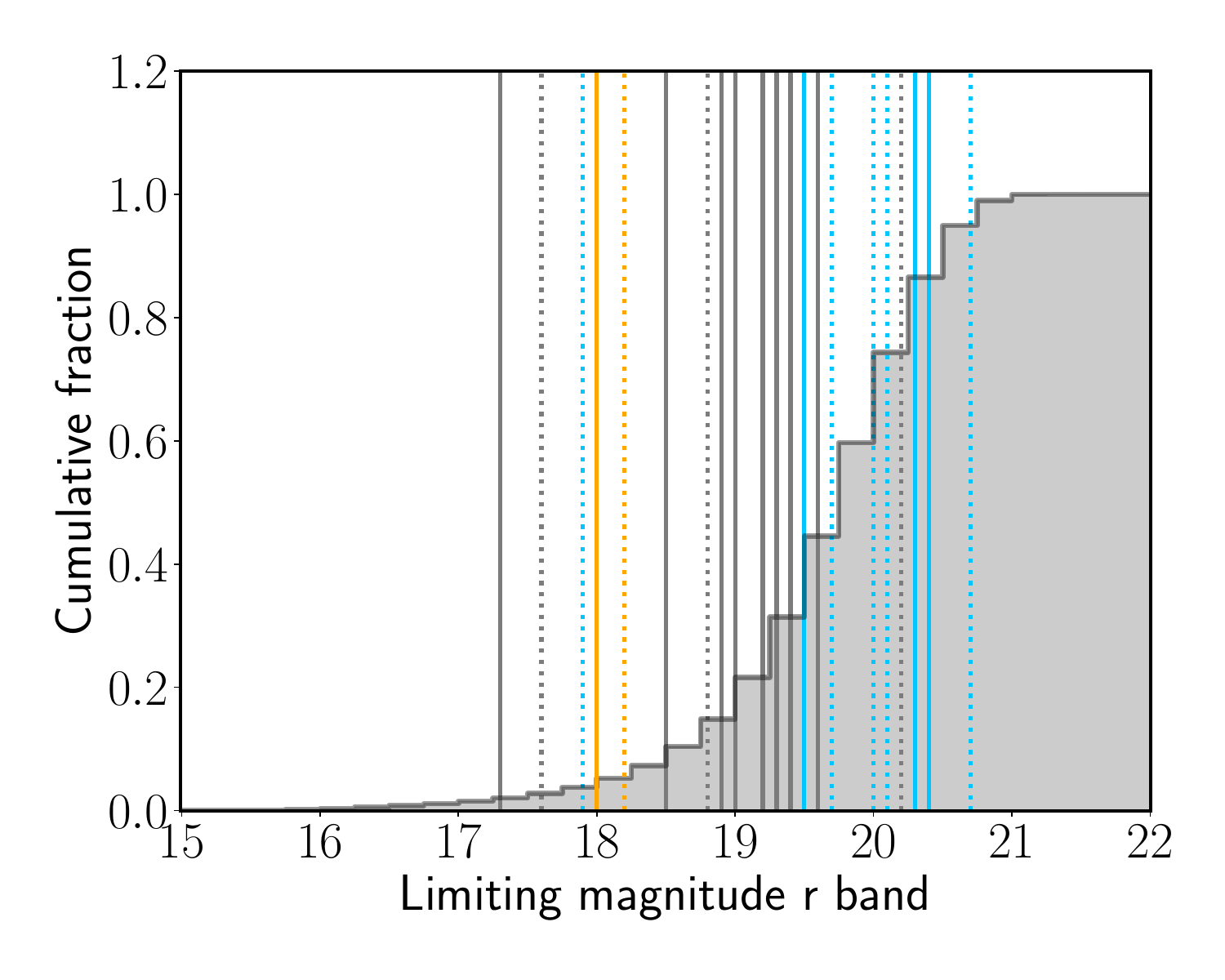} 
\caption{\textbf{Sun-centric angular distance and 5-$\sigma$ limiting magnitude distributions of twilight survey fields taken between 2019 September and 2022 September.} \textbf{Top left panel:} the Sun-centric angular distance distribution of evening twilight survey fields taken between 2021 April and 2022 March for the evening portion and 2021 April and 2022 September for the morning potion. \textbf{Top right panel:} the cummulative Sun-centric angular distance distribution of evening twilight survey fields taken between 2021 April and 2022 March for the evening portion and 2021 April and 2022 September for the morning potion. \textbf{Center left panel:} the cumulative Sun-centric angular distance distribution of evening twilight survey fields taken between 2019 September and 2022 March. \textbf{Center right panel:} the cumulative Sun-centric angular distance distribution of morning twilight survey fields taken between 2019 September and 2022 September. \textbf{Bottom left panel:} the cumulative Sun-centric angular distance distribution of evening and morning twilight survey fields taken between 2019 September and 2022 March for the evening portion and 2019 September and 2022 September for the morning portion. \textbf{Bottom right panel:} the cumulative 5-$\sigma$ limiting r magnitude distribution of evening and morning twilight survey fields taken between 2019 September and 2022 March for the evening portion and 2019 September and 2022 September for the morning portion. Orange lines are indicated for \an discovery and recoveries. Blue lines are indicated for Atira discovery and recoveries. Both long and short-period comets are indicated by grey lines. The lines will be solid for ZTF discoveries and dashed for recoveries.}
\end{figure}

The 5-$\sigma$ limiting r-magnitude for the evening portion between 2019 September and 2022 March and for the morning portion between 2019 September and 2022 September are shown in the bottom right panel of Fig.~3. The median  5-$\sigma$ limiting r-magnitude was $\sim$19.9 during this period. The limiting magnitude of twilight survey fields varied significantly with the time after/before sunset, the altitude of the Sun at the time the fields were taken, and the angular distance between the area of sky covered by the survey and the Sun. As seen in the top, center, and bottom panels of Fig.~4-6, the limiting magnitude generally grows fainter with longer time after/before sunset, smaller Sun altitude, and more significant angular distance for both evening and morning twilight surveys. As seen in the top, center, and bottom panels of Fig.~6, the limiting magnitude decreases well below 20 mags for Solar angular distances smaller than 46$^{\circ}$ where \an objects can be found \citep[][]{Bolin2022IVO}.

\begin{figure}\centering
\includegraphics[width=0.5\linewidth]{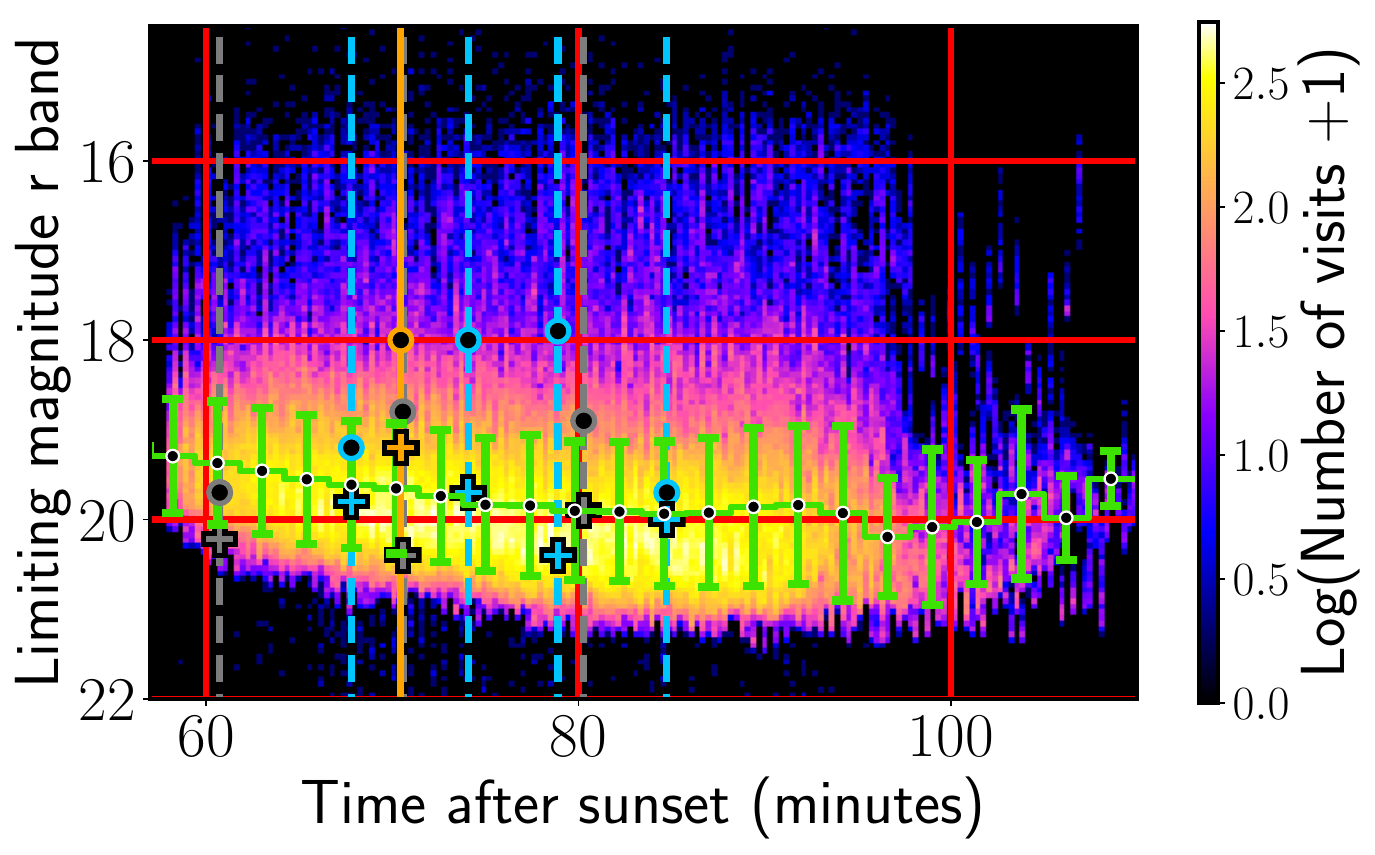} 
\includegraphics[width=0.5\linewidth]{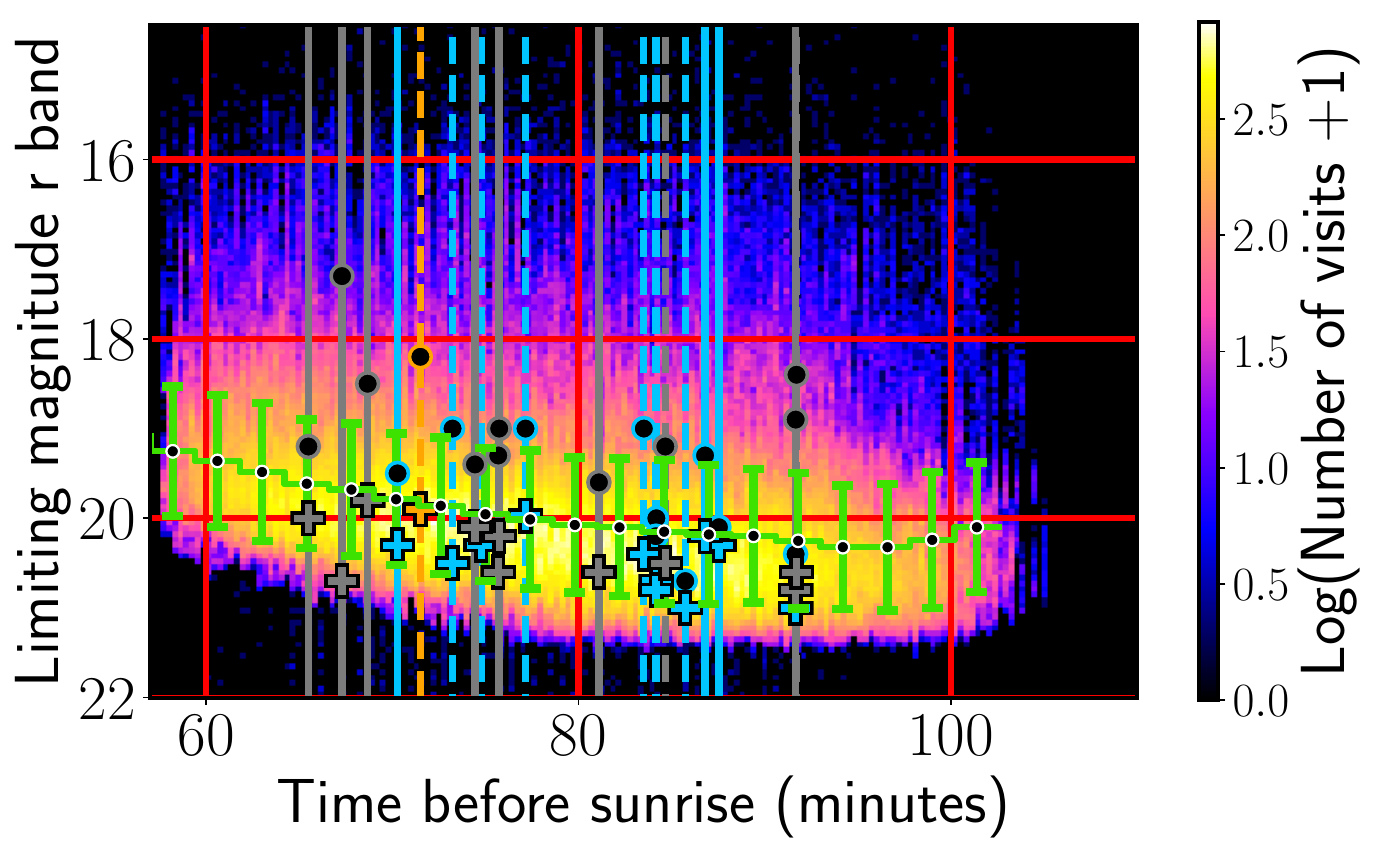}
\includegraphics[width=0.5\linewidth]{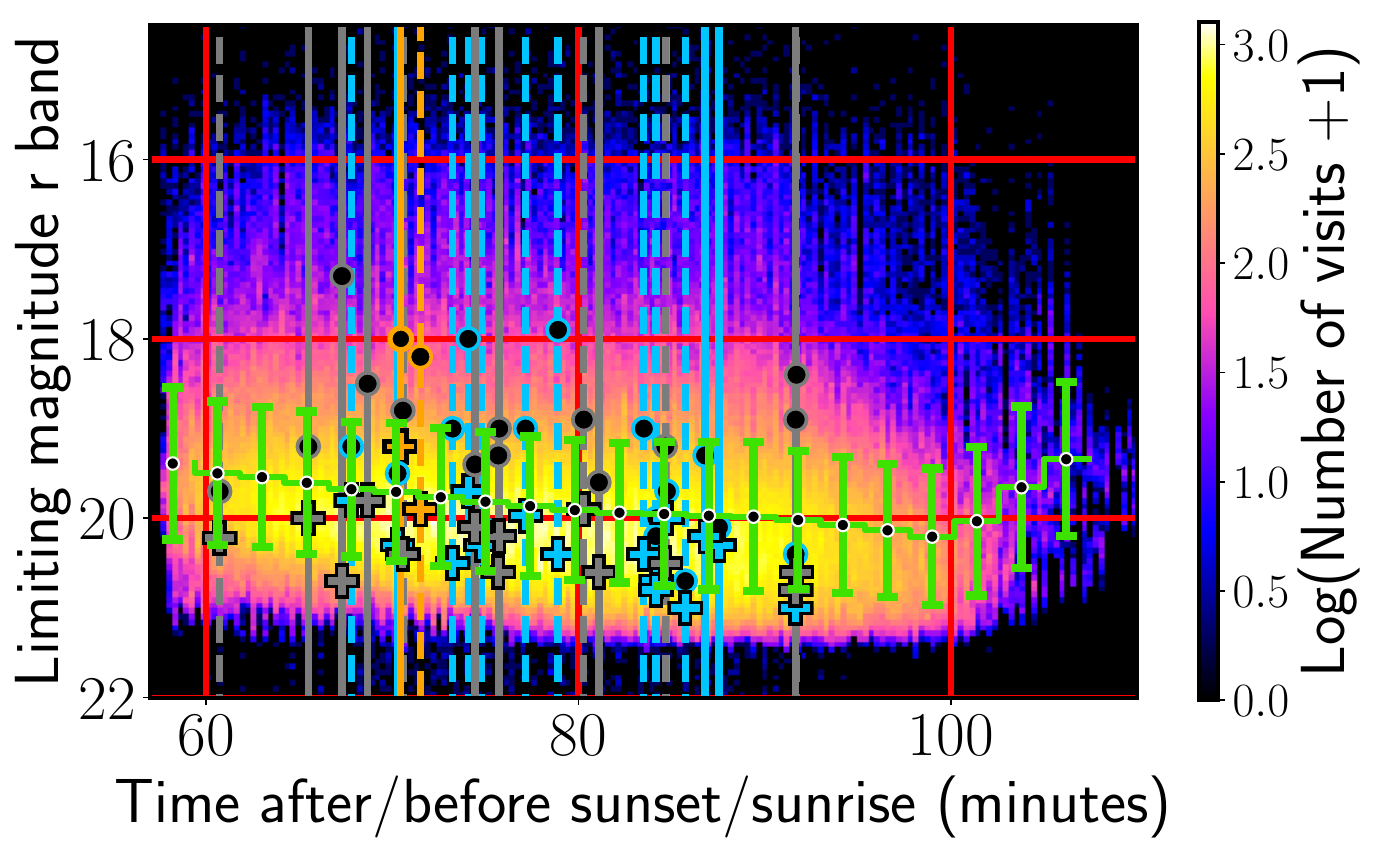} 
\caption{\textbf{Time after sunset/before sunrise and limiting magnitude distributions of twilight survey fields taken between 2019 September and 2022 September.} \textbf{Top panel:} the time after sunset vs limiting magnitude distribution of evening twilight survey fields taken between 2019 September and 2022 March. \textbf{Middle panel:} the time before sunset vs limiting magnitude distribution of evening twilight survey fields taken between 2019 September and 2022 September. \textbf{Bottom panel:} the time after sunset vs limiting magnitude distribution of evening twilight survey fields taken between 2019 September and 2022 March and time before sunrise 2019 September and 2022 September for the morning portion. Orange lines are indicated for \an discovery and recoveries. Blue lines are indicated for Atira discovery and recoveries. Grey lines indicate both long and short-period comets. The lines will be solid for discoveries and dashed for recoveries. The filled-in plus symbols show the limiting magnitude of the image in which a discovery or recovery was made. The filled-in black circle shows the magnitude of the asteroid or comet discovery or recovery. The bin size after sunset/before sunrise was 0.3 minutes, and the bin size in the magnitude direction was 0.06125 mags. The color scale indicates the Log number of visits per 2D bin +1. The green error bar line indicates the running mean of the limiting magnitude in increments of 2.4 minutes after/before Sunset with error bars corresponding to the per bin 1-$\sigma$ standard deviation value.}
\end{figure}

\begin{figure}\centering
\includegraphics[width=0.5\linewidth]{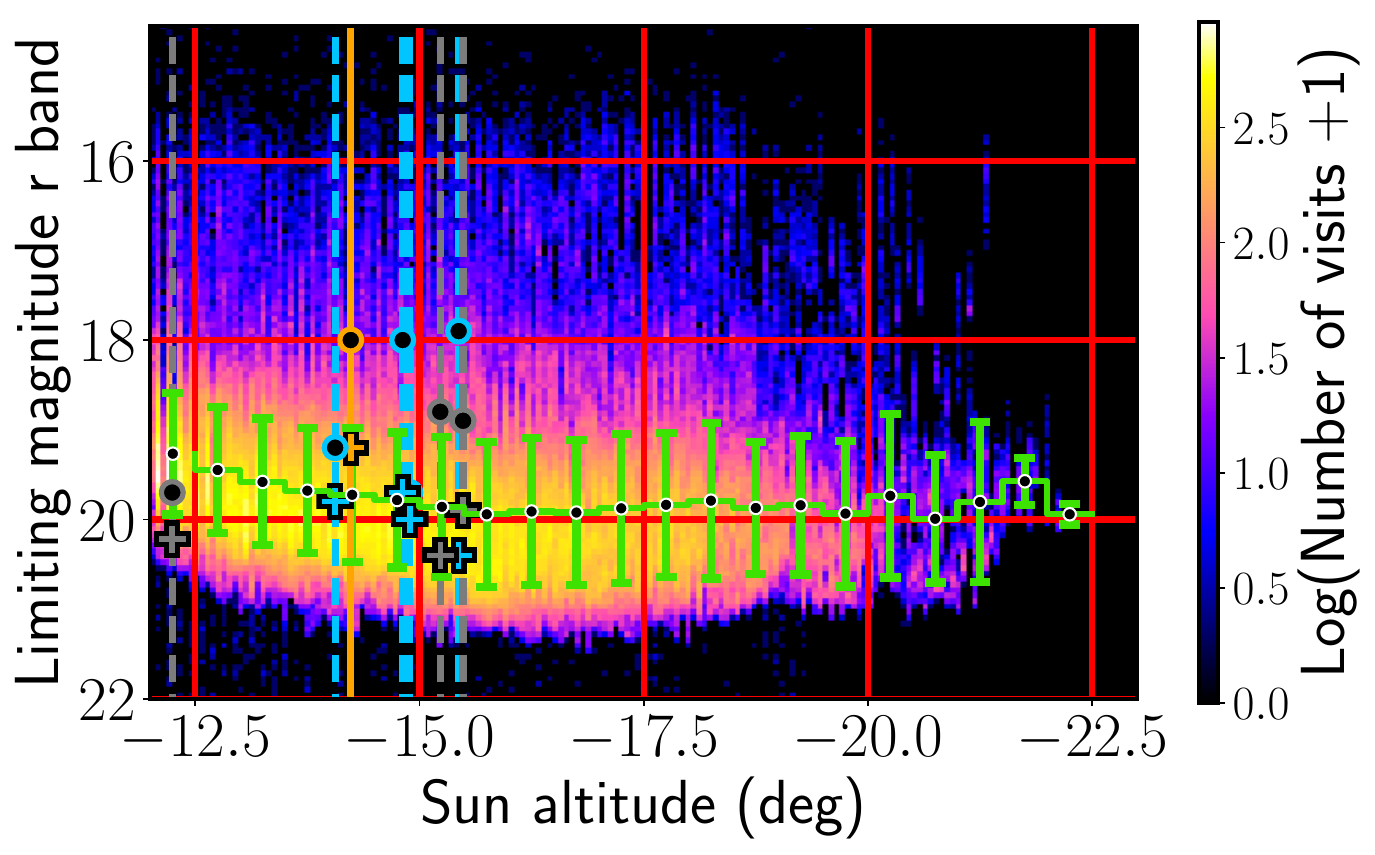} 
\includegraphics[width=0.5\linewidth]{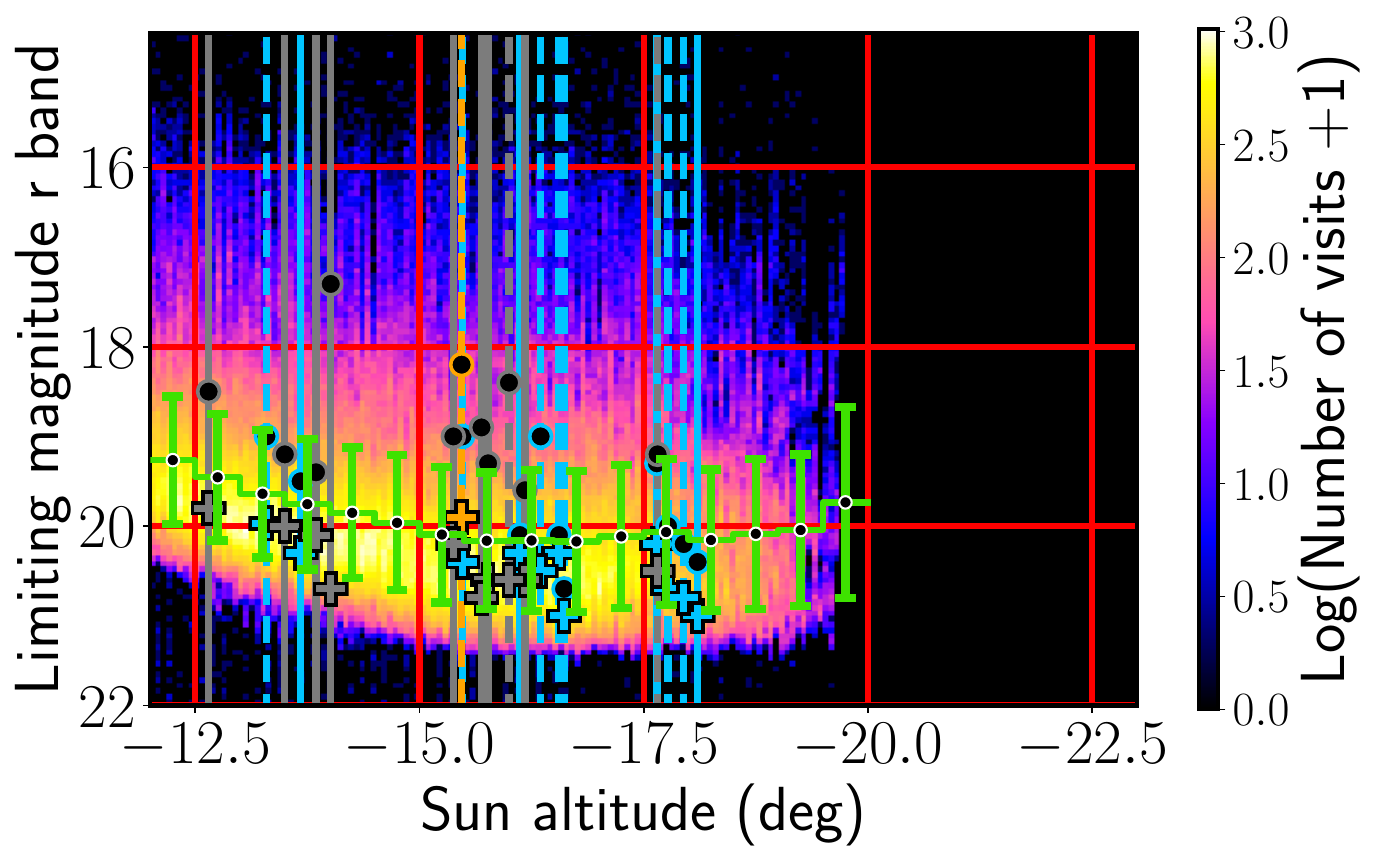}
\includegraphics[width=0.5\linewidth]{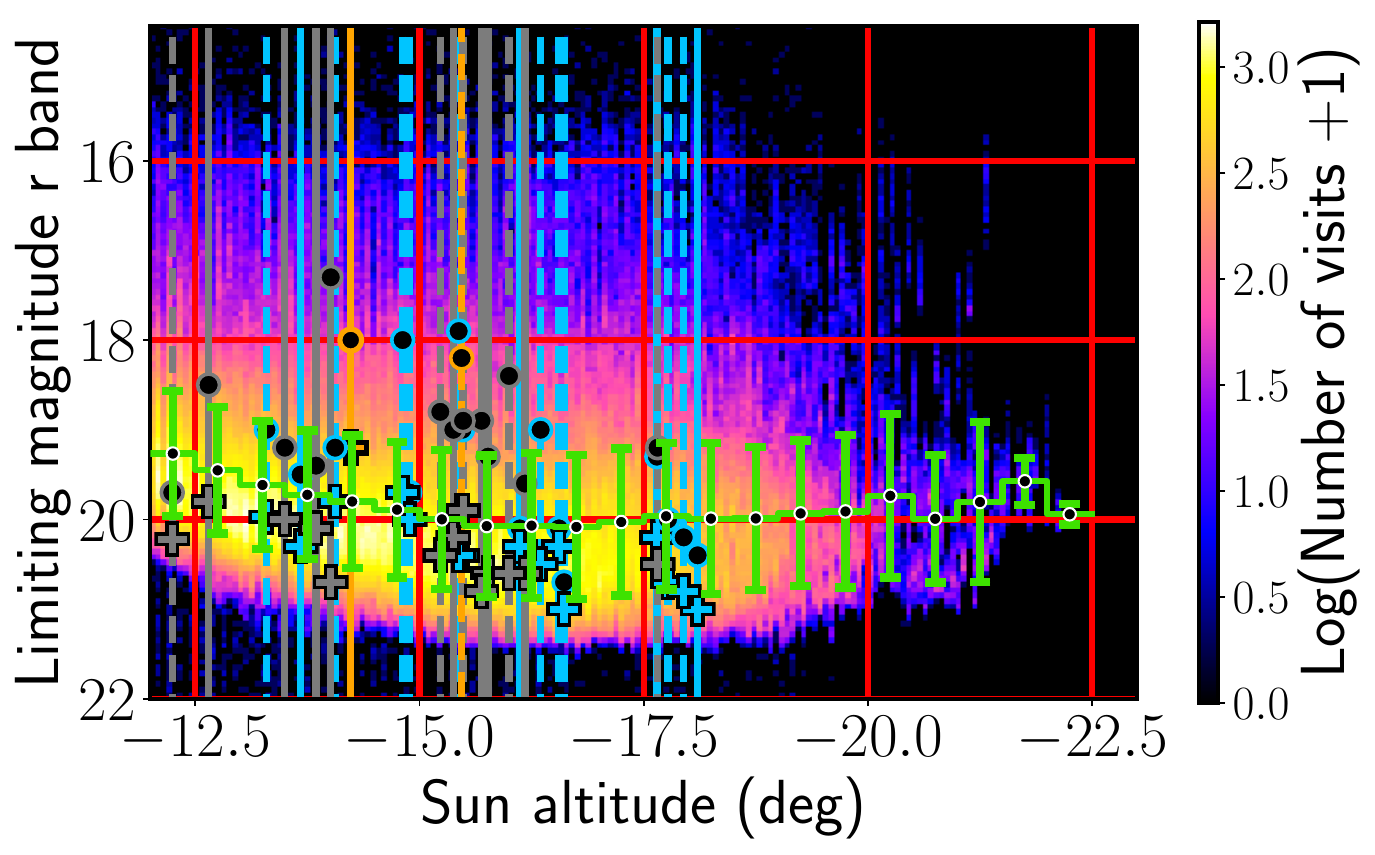} 
\caption{\textbf{Sun altitude and limiting magnitude distributions of twilight survey fields taken between 2019 September and 2022 September.} \textbf{Top panel:} the Sun altitude vs limiting magnitude distribution of evening twilight survey fields taken between 2019 September and 2022 March. \textbf{Middle panel:} the Sun altitude vs limiting magnitude distribution of evening twilight survey fields taken between 2019 September and 2022 September. \textbf{Bottom panel:} the Sun altitude vs limiting magnitude distribution of evening twilight survey fields taken between 2019 September and 2022 March for the evening portion and 2019 September and 2022 September for the morning portion. Orange lines are indicated for \an discovery and recoveries. Blue lines are indicated for Atira discovery and recoveries. Both long and short-period comets are indicated by grey lines. The lines will be solid for discoveries and dashed for recoveries. Orange lines are indicated for \an discovery and recoveries. Blue lines are indicated for Atira discovery and recoveries. Grey lines indicate both long and short-period comets.  The lines will be solid for ZTF discoveries and dashed for recoveries. The filled-in plus symbols show the limiting magnitude of the image in which a discovery or recovery was made. The filled-in black circle shows the magnitude of the asteroid or comet discovery or recovery. The bin size in the Sun altitude direction is 0.06125$^{\circ}$, and the bin size in the magnitude direction was 0.06125 mags. The color scale indicates the Log number of visits per 2D bin +1. The green error bar line indicates the running mean of the limiting magnitude in increments of 0.5$^{\circ}$ Sun altitude with error bars corresponding to the per bin 1-$\sigma$ standard deviation value.}
\end{figure}

\begin{figure}\centering
\includegraphics[width=0.5\linewidth]{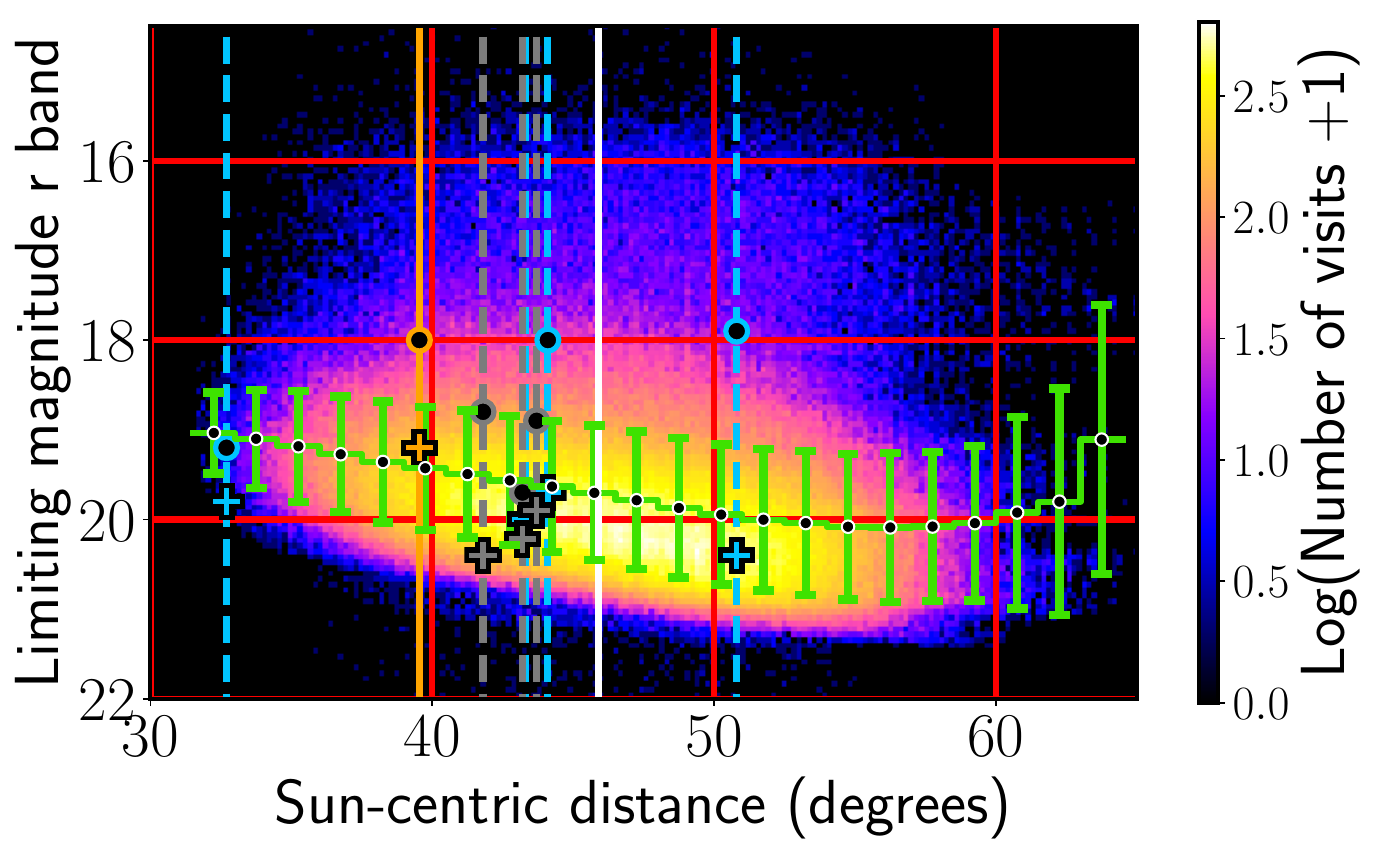} 
\includegraphics[width=0.5\linewidth]{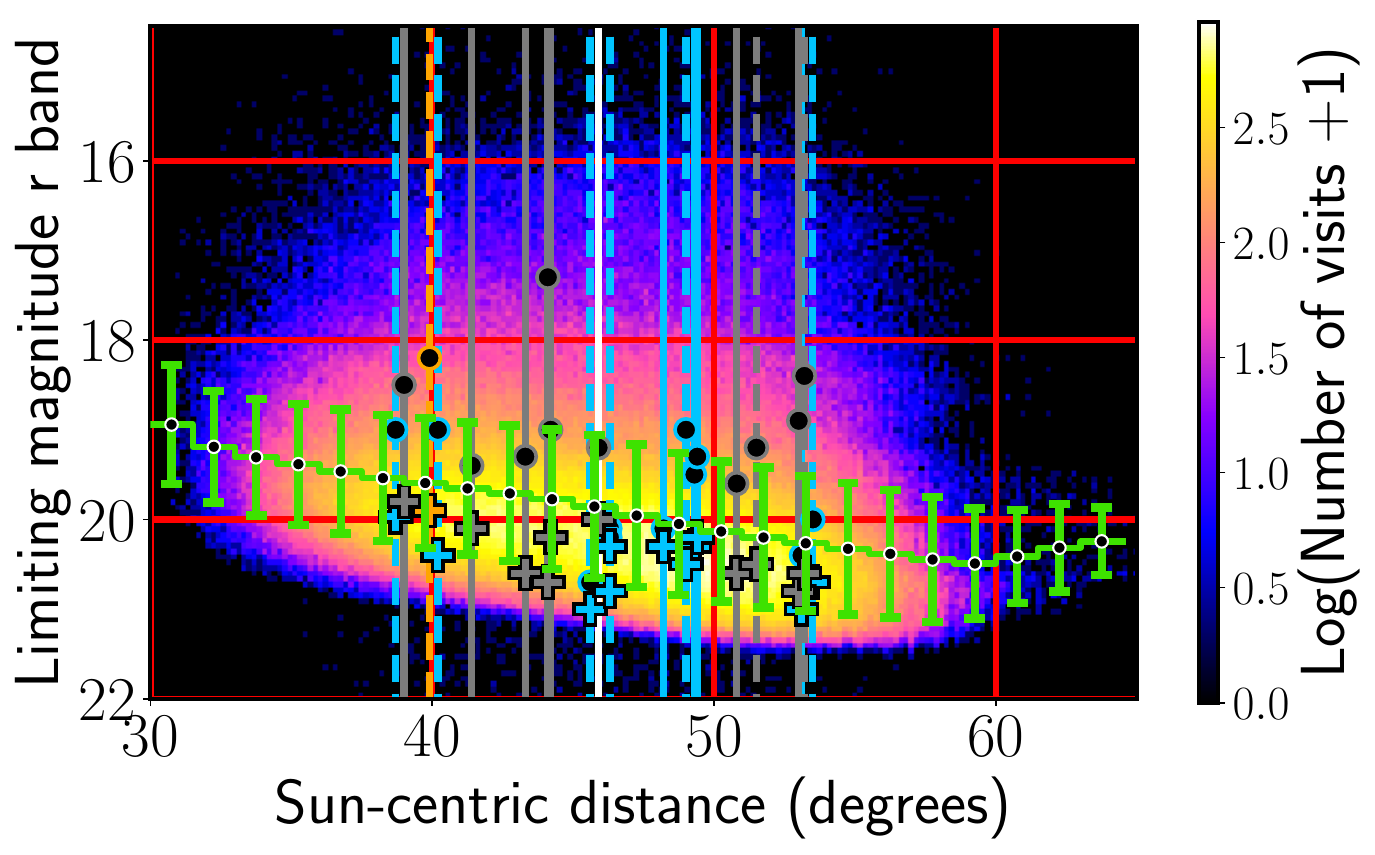}
\includegraphics[width=0.5\linewidth]{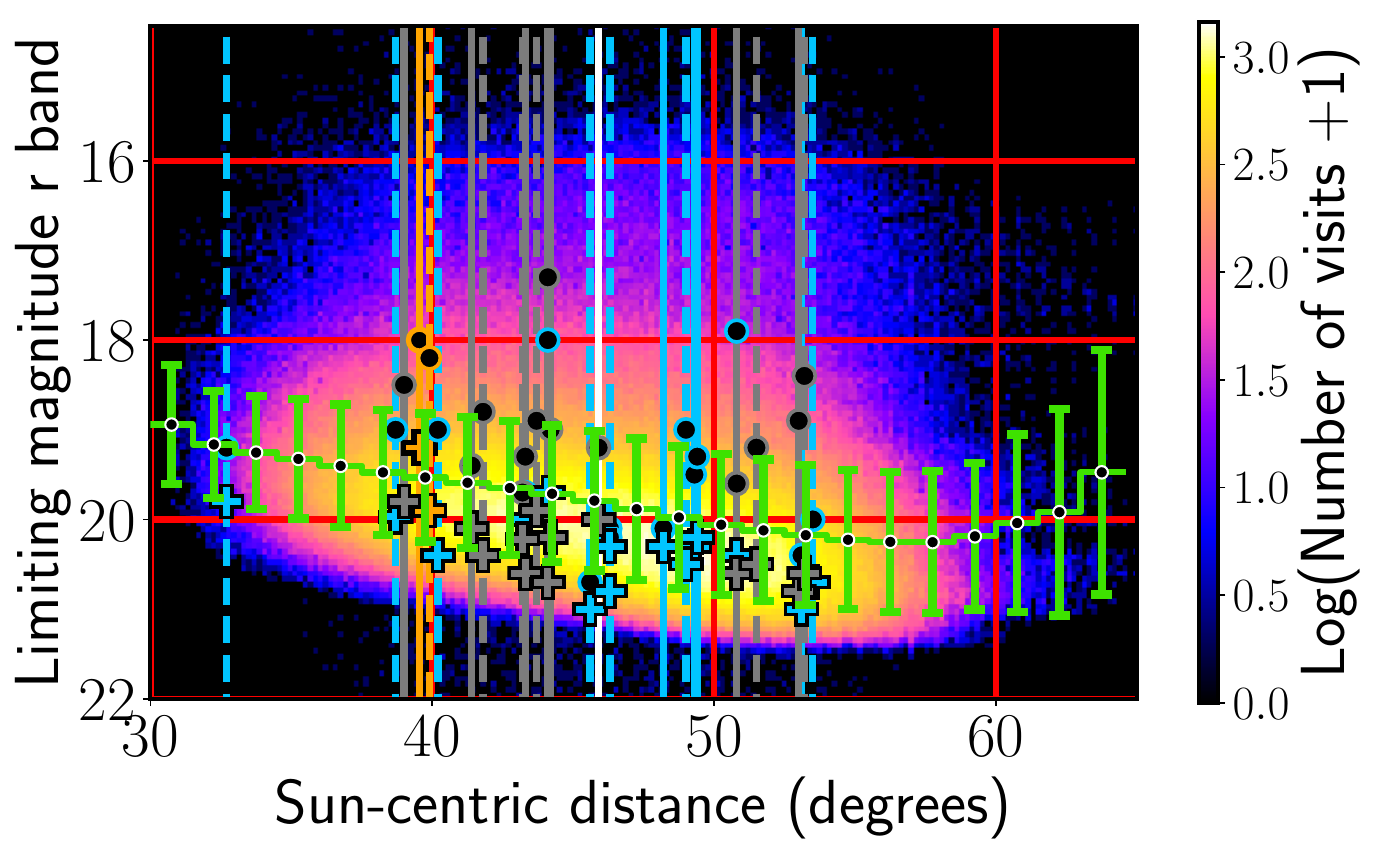} 
\caption{\textbf{Solar angular distance and limiting magnitude distributions of twilight survey fields taken between 2019 September and 2022 September.} \textbf{Top panel:} the Solar angular distance vs limiting magnitude distribution of evening twilight survey fields taken between 2019 September and 2022 March. \textbf{Middle panel:} the Solar angular distance vs limiting magnitude distribution of evening twilight survey fields taken between 2019 September and 2022 September. \textbf{Bottom panel:} the Solar angular distance vs limiting magnitude distribution of evening twilight survey fields taken between 2019 September and 2022 March for the evening portion and 2019 September and 2022 September for the morning portion. Orange lines are indicated for \an discovery and recoveries. Blue lines are indicated for Atira discovery and recoveries. Grey lines indicate both long and short-period comets. Solid lines are used for discoveries and dashed for recoveries. The white line is the maximum angular distance of \an objects at 46$^{\circ}$ from the Sun. Orange lines are indicated for \an discovery and recoveries. The filled-in plus symbols show the limiting magnitude of the image in which a discovery or recovery was made. The filled-in black circle shows the magnitude of the asteroid or comet discovery or recovery. The bin size in the Sun altitude direction is 0.18$^{\circ}$, and the bin size in the magnitude direction was 0.06125 mags. The color scale indicates the Log number of visits per 2D bin +1. The green error bar line indicates the running mean of the limiting magnitude in increments of 1.5$^{\circ}$ of Solar angular distance with error bars corresponding to the per bin 1-$\sigma$ standard deviation value. }
\end{figure}

Slight seasonal variations were also present in the solar angular distance coverage and limiting magnitude of the twilight survey. The top left panel, top right, and bottom panels of Fig.~7 show the day of the year vs. Sun-centric angular distance distributions for the evening morning and combined evening and morning twilight surveys. The evening twilight survey showed a small increasing trend in Sun-centric angular distance with a day of the year. An average Sun-centric angular distance value of $\sim$48$^{\circ}$ for fields taken past 300 days of the year compared to $\sim$46$^{\circ}$ at the start of the year. This is likely due to seasonal variations in the amount of inclement weather causing the survey to start at later times in the evening. The top left, top right, and bottom panels of Fig.~8 show the mean limiting r-band magnitude as a function of the day of the year and Sun-centric angular distance. The overall trend of worsening limiting magnitude with decreasing Sun-centric angular distance shown in Fig.~6 is similarly seen in Fig.~8. Seasonal variations are present in the limiting magnitude, with there being slightly higher limiting r-magnitudes of $\sim$20.2 on average in the Summer compared to an average of $\sim$19.8 r-magnitude in the winter for fields more than 50$^{\circ}$ from the Sun.

\begin{figure}\centering
\includegraphics[width=0.49\linewidth]{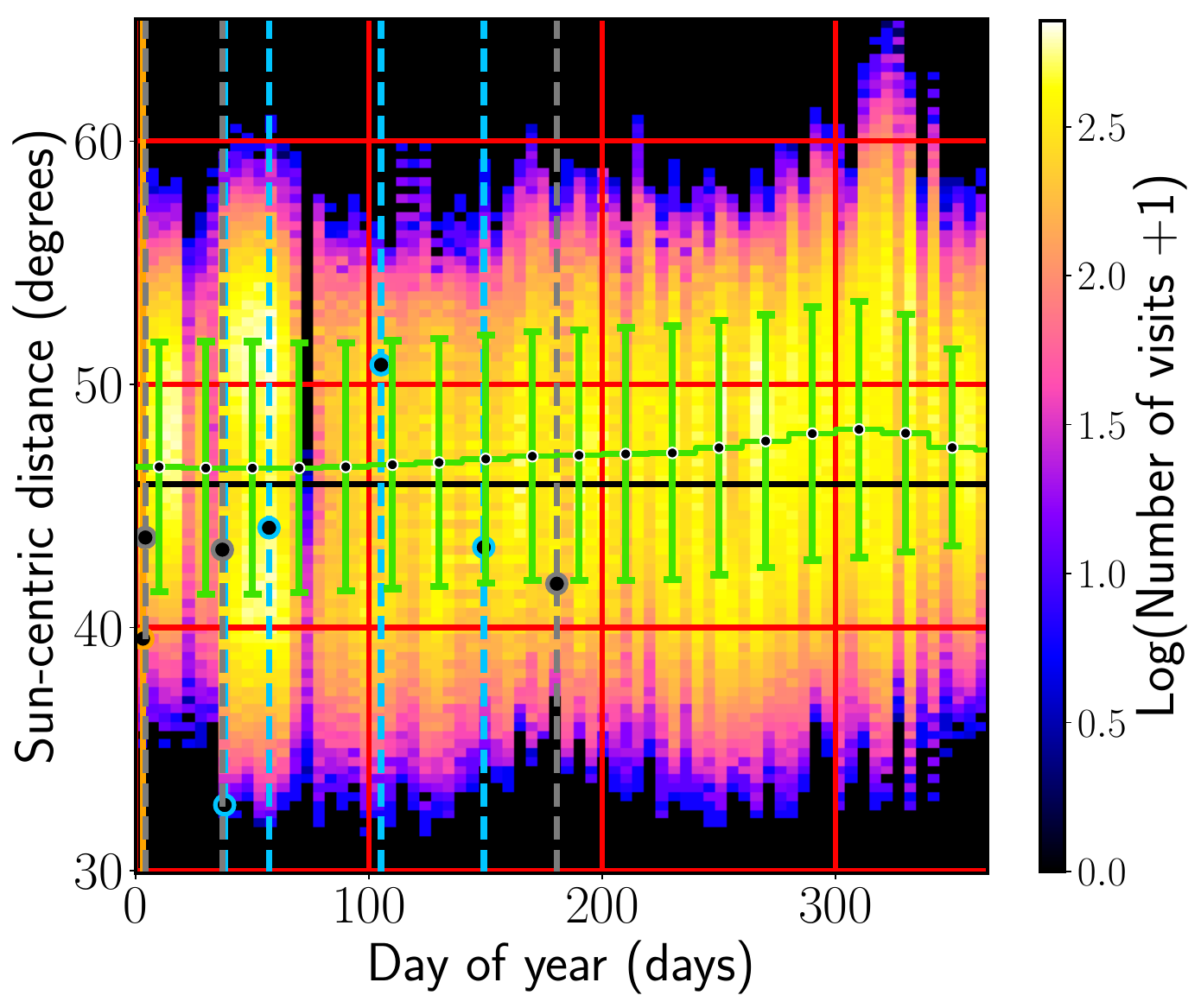} 
\includegraphics[width=0.49\linewidth]{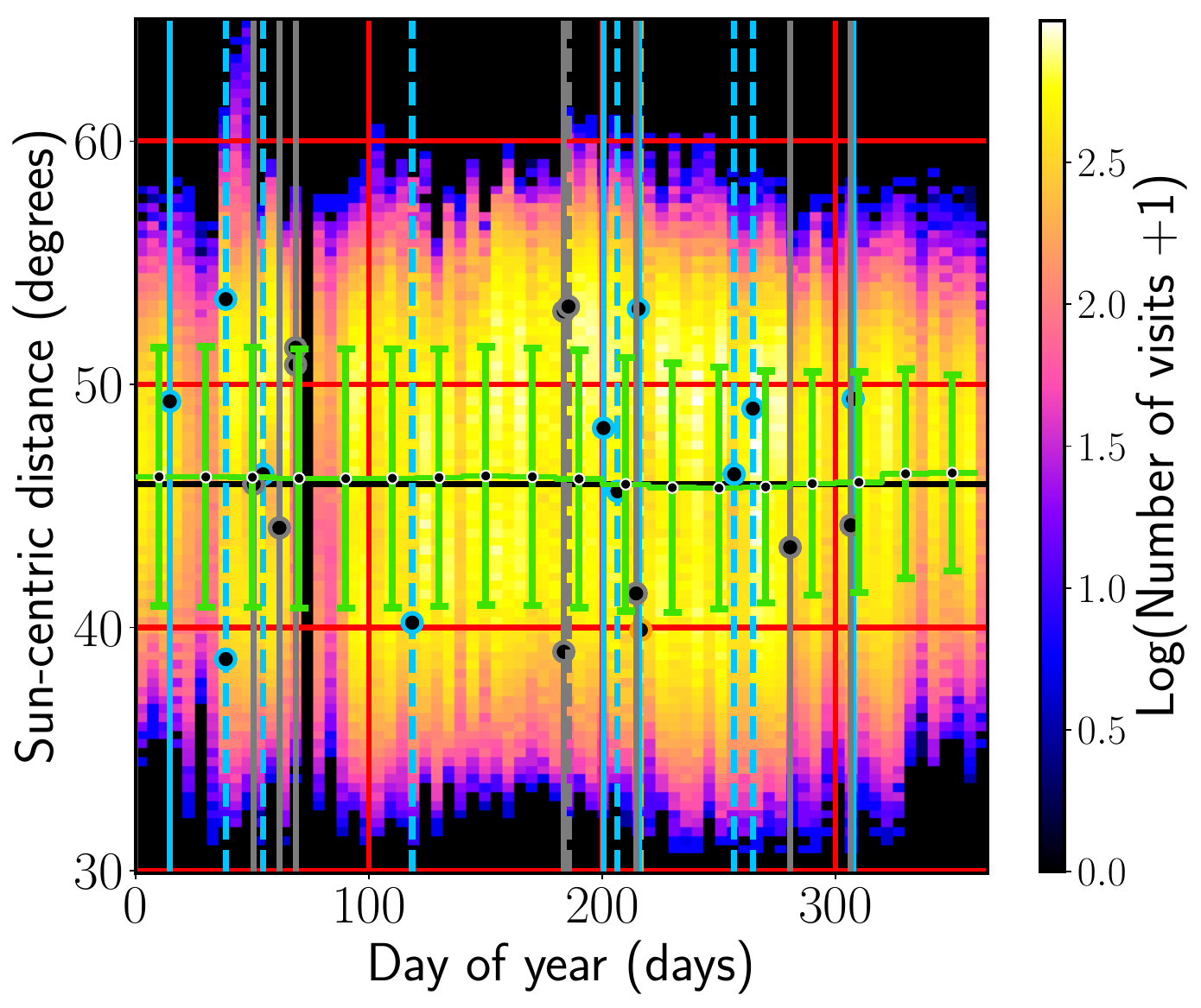}
\includegraphics[width=0.49\linewidth]{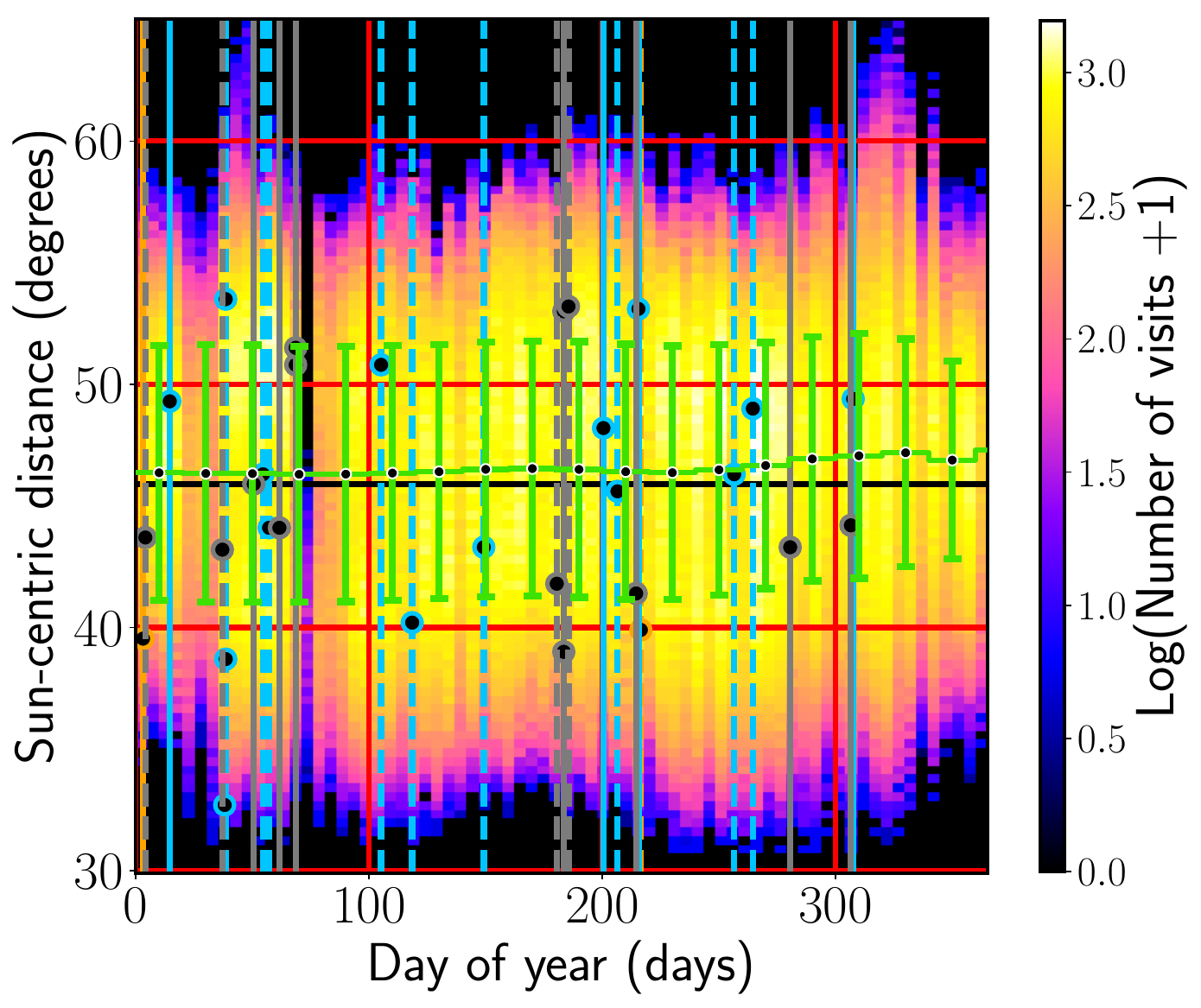} 
\caption{\textbf{Day of year and Sun-centric angular distance distributions of twilight survey fields taken between 2019 September and 2022 September.} \textbf{Top left panel:} the day of year vs Sun-centric angular distance distribution of evening twilight survey fields taken between 2019 September and 2022 March. \textbf{Top right panel:} the day of the year vs. Sun-centric angular distance distribution of morning twilight survey fields taken between 2019 September and 2022 September. \textbf{Bottom panel:} the day of year vs. Sun-centric angular distance distribution of evening twilight survey fields taken between 2019 September and 2022 March for the evening portion and 2019 September and 2022 September for the morning portion. Orange lines are indicated for \an discovery and recoveries. Blue lines are indicated for Atira discovery and recoveries. Grey lines indicate both long and short-period comets. The lines will be solid for ZTF discoveries and dashed for recoveries. The black line is the maximum angular distance of \an objects at 46$^{\circ}$ from the Sun. The filled-in black circle shows the Sun-centric angular distance of the asteroid or comet discovery or recovery. The bin size in the day-of-year direction is five days, and the bin size in the Sun-centric angular distance direction is 0.36$^{\circ}$. The color scale indicates the Log number of visits per 2D bin +1. The black horizontal line is the maximum angular distance of \an objects at 46$^{\circ}$ from the Sun. The green error bar line indicates the running mean of the Sun-centric angular distance in increments of 20 days with error bars corresponding to the per bin 1-$\sigma$ standard deviation value.}
\end{figure}

\begin{figure}\centering
\includegraphics[width=0.49\linewidth]{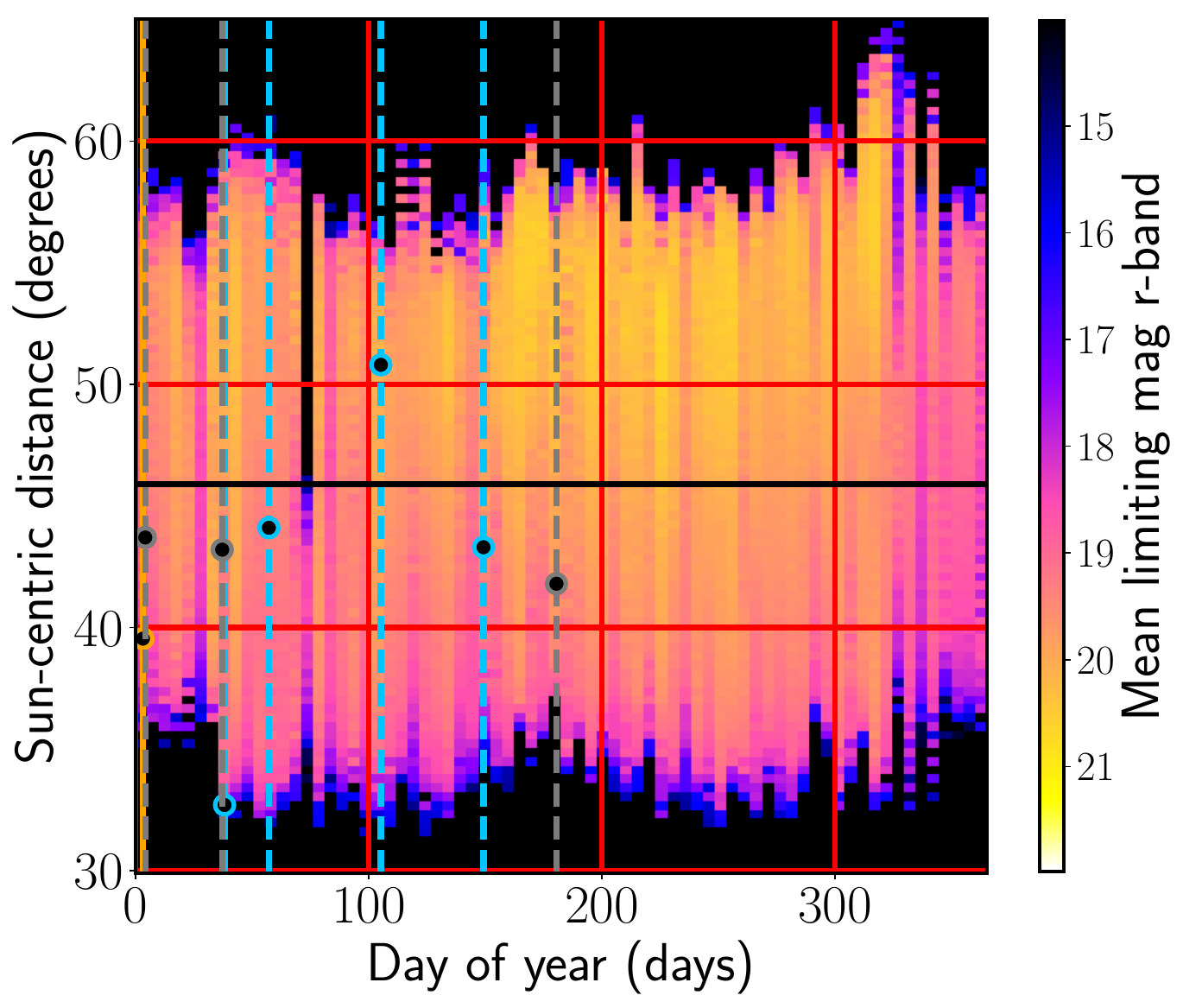} 
\includegraphics[width=0.49\linewidth]{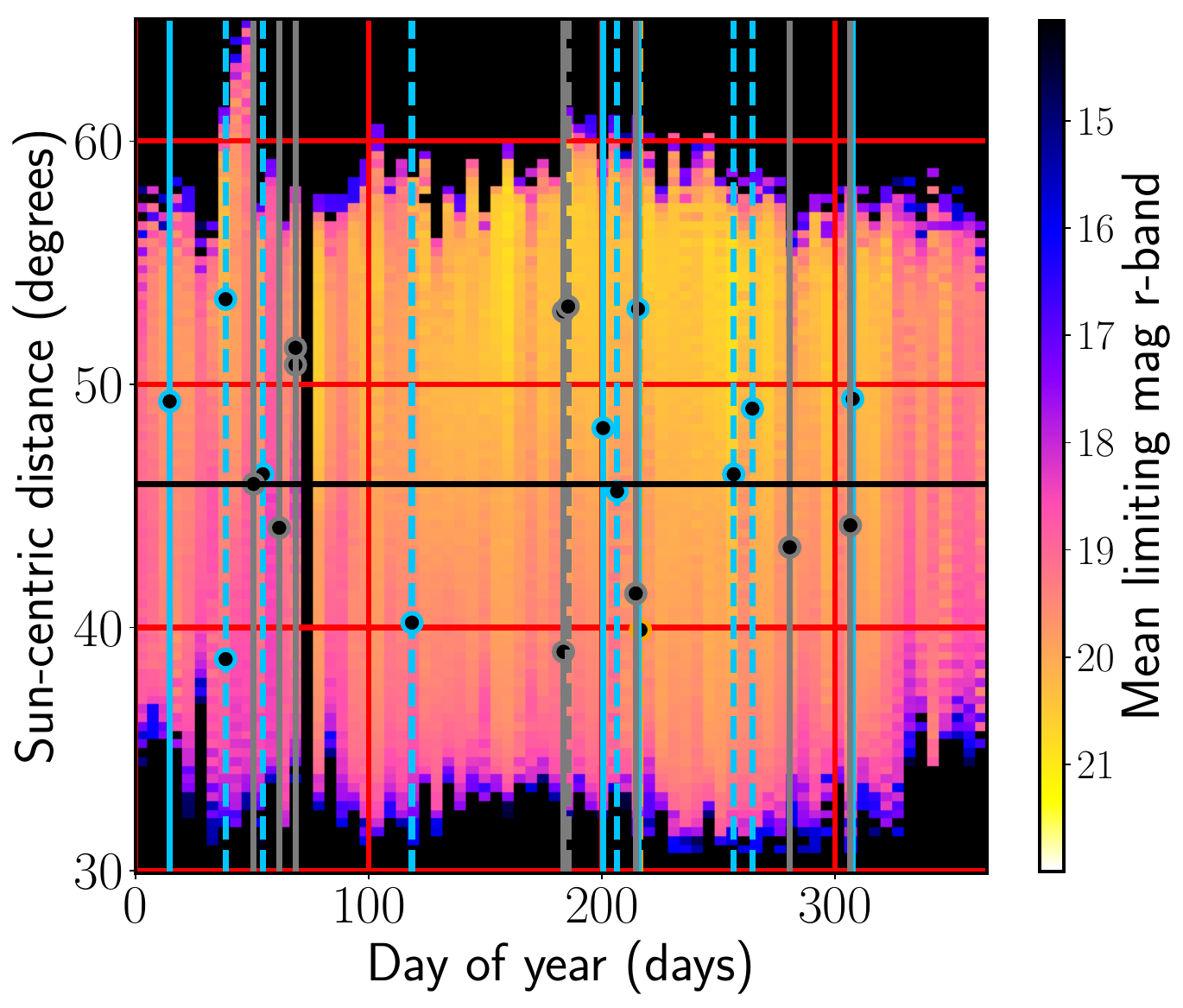}
\includegraphics[width=0.49\linewidth]{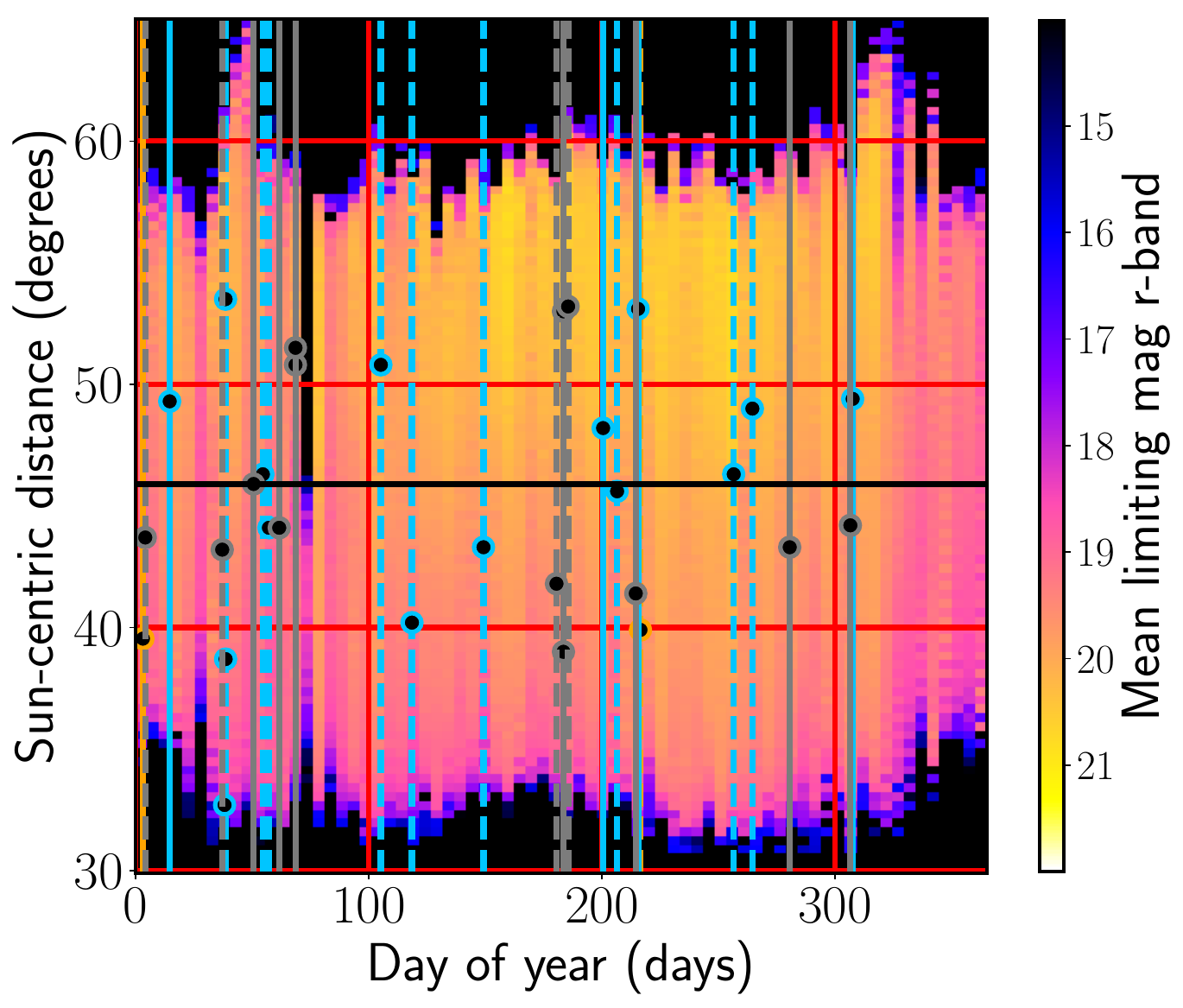} 
\caption{\textbf{Solar angular distance and limiting magnitude distributions of twilight survey fields taken between 2019 September and 2022 September as a function of day of the year.} \textbf{Top left panel:} the Solar angular distance vs limiting magnitude distribution of evening twilight survey fields taken between 2019 September and 2022 March. \textbf{Top right panel:} the Solar angular distance vs limiting magnitude distribution of morning twilight survey fields taken between 2019 September and 2022 September. \textbf{Bottom panel:} the Solar angular distance vs limiting magnitude distribution of evening twilight survey fields taken between 2019 September and 2022 March for the evening portion and 2019 September and 2022 September for the morning portion. Orange lines are indicated for \an discovery and recoveries. Blue lines are indicated for Atira discovery and recoveries. Both long and short-period comets are indicated by grey lines.  The lines will be solid for ZTF discoveries and dashed for recoveries. The black line is the maximum angular distance of \an objects at 46$^{\circ}$ from the Sun. The filled-in black circle shows the Sun-centric angular distance of the asteroid or comet discovery or recovery. The bin size in the Sun altitude direction is 0.18$^{\circ}$, and the bin size in the magnitude direction was 0.06125 mags. The color scale indicates the Log number of visits per 2D bin +1. The green error bar line indicates the running mean of the limiting magnitude in increments of 1.5$^{\circ}$ of Solar angular distance.}
\end{figure}

\subsection{Follow up observation details}

The initial observations of asteroid and comet discovery candidates were reported to the Minor Planet Center (MPC) within several hours of the twilight survey sessions for ZMODE Lite asteroid candidates, and within 5-10 h for Tails comet candidates. A minimum of three observations were reported to MPC. Additionally, announcement of the discovery of candidates were also announced to the Global Relay of Observatories Watching Transients Happen network (GROWTH) \cite{Kasliwal2019}. The GROWTH network was established to follow up supernova and gravitational wave events but has been previously used to follow up solar system objects detected by ZTF for both characterization and astrometry \citep[][]{Bolin20202I,Bolin2021LD2,Purdum2021,Farnocchia2022, Sharma2023}. The few hours difference between the observations and the reporting of candidates to the Minor Planet Center and GROWTH network and the $\sim$10-20 minutes length of observational arc typical for twilight observations resulted in our candidates having relatively small skyplane uncertainties of $\sim$10 arcsec and growing by $\sim$1 arcmin per day\footnote{As indicated for our candidates' entries on the MPC's near-Earth object confirmation page: \url{https://minorplanetcenter.net//iau/NEO/toconfirm_tabular.html}}. Astrometry was measured in the images taken by the follow up observatories and submitted to the MPC. We will describe the various follow up assets used during the twilight survey below. The typical observational arc for the combination of ZTF and GROWTH follow up observations was typically $\sim$1-2 days.

\textit{GROWTH-India}
The 0.7-m GROWTH-India telescope, located at the Indian Astronomical Observatory (IAO), located atop Mt. Saraswati, Digpa Ratsa Ri in Hanle, Ladakh, (observatory code L51) was used to follow up twilight survey candidates. Follow up observations were completed using the SDSS r-band filter and Andor iKon-XL 230 CCD camera with a 0.7$^{\circ}$  wide field of view and 0.676 arcsec pixel scale \citep[][]{Kumar2022,Sharma2023}. Non-sidereal tracking at the rate of the targets was used. Data were detrended, and astrometry was measured in real-time at IIT Bombay. Astrometric solutions were obtained using the offline engine of astrometry.net \citep{Lang2010}. L51 was used to follow up C/2020 T2 on 2020-10-21, C/2020 V2 on 2020-11-07, 2020-11-11, and 2020-11-12, C/2021 D2 on 2021-03-11, C/2021 E3 on 2021-03-14, 2021-04-14, 2021-04-18, P/2021 N1 on 2021-07-07.

\textit{Mount Laguna Observatory 40-inch Telescope}
Astrometric follow up of ZTF twilight discoveries were made using the 1.0 m Telescope at the Mount Laguna Observatory (observatory code U83) \citep[][]{SmithNelson1969}. Astrometric follow up was made using the $R$ filter in combination with the E2V 42-40 CCD Camera with a field of view of 12~arcmin x 12~arcmin and pixel scale of 0.36~arcsec/pixel. Non-sidereal tracking was used in the observation of the twilight survey targets at their predicted rate of motion. U83 was used to follow up (594913) \an on 2021-08-12 and 2021-08-23, C/2021 E3 on 2021-04-01, P/2021 N1 on 2021-07-08.

\textit{Liverpool Telescope}
The 2-m Liverpool telescope is located at the Observatorio del Roque de los Muchachos (observatory code J13). Follow up observations of ZTF twilight discoveries were made using the IO:O wide-field camera with a 10 arcminute x 10 arcminute wide field of view using the SDSS r-band filter. Follow up targets were tracked at a non-sidereal rate set by their predicted rate of motion in the night sky. The camera was used with a 2x2 binning, providing a pixel scale of 0.3 arcsec~ and data were detrended using the pipeline software \citep[][]{Steele2004}. J13 was used to follow up 2020 OV$_1$ on 2020-07-21 and 2020-07-30, P/2021 N1 on 2021-07-08, 2021-07-09, 2021-07-10, 2021-07-11.

\textit{Lulin Optical Telescope}
The 1 m Lulin Optical Telescope (observatory code D35) was used to follow up on the twilight survey targets. Astrometric observations were taken using the 2K $\times$ 2K SOPHIA camera with a field of view of 17~arcmin x 17~arcmin and a pixel scale of 0.52  arcsec/pixel \citep[][]{Kinoshita2005}. The follow up targets were tracked at their predicted rate of motion using non-sidereal tracking. D35 was used to follow up (594913) \an on 2020-11-26

\textit{Kitt Peak Electron Multiplying CCD Demonstrator (KPED):} The KPED instrument mounted on the Kitt Peak 84-inch telescope (observatory code 695) was used to obtain astrometric follow up observations of twilight survey targets. The camera consists of a 1024 $\times$ 1024 pixel Electron Multiplying CCD camera with a pixel scale of 0.26 arcseconds pixel$^{-1}$ and a 4.4~arcminute $\times$ 4.4~arcminute field of view. Twilight survey targets were non-sidereally tracked at their apparent sky motion rates, and the SDSS r-band filter was used. The Kitt Peak 84-inch telescope was robotically operated \citep[][]{Coughlin2019}. 695 was used to follow up (594913) \an on 2021-08-12 and 2021-08-23.

\textit{Magellan Telescope:} The 6.5~m Magellan Baade telescope at Las Campanas Observatory (observatory code 269) was used to follow up twilight survey targets with the FourStarr near-infrared camera with a 10.9~arcmin x 10.9~arcmin field of view and 0.16~arcsec/pixel image scale \citep[][]{Persson2013}. Data were taken in the J band filter while the telescope was tracked non-sidereally at the target's on-sky rate of motion. 269 was used to follow up (594913) \an on 2021-07-18 and 2021-07-19.

\textit{Spectral Energy Distribution Machine (SEDM):} Follow up for the twilight survey was made using the Rainbow Camera of the SEDM (observatory code 675) \citep[][]{Blagorodnova2018}. The Rainbow Camera consists of two identical Princeton Instruments Pixis 2048B eXelon model 2048~pixel $\times$ 2048~pixel CCDs mounted on the Palomar 60-inch telescope and is robotically operated \citep[][]{Blagorodnova2018,Rigault2019}. The Rainbow Camera field of view consists of four $\sim$6 arcminute quadrants, with each having coverage of a single SDSS filter, u, g, r, and i at a 0.37 arcseconds pixel$^{-1}$ spatial scale providing a total 13 arcminute field of view. The follow up targets were tracked non-sidereally and placed into the r-band quadrant. 675 was used to follow up (594913) \an on 2020-01-08 and 2020-11-24, 2020 OV$_1$ on 2020-07-30 and 2020-08-06, C/2020 T2 on 2020-10-11, C/2020 V2 on 2020-11-17.

\textit{Southern Astrophysical Research Telescope (SOAR):} 
The 4.1-m SOAR telescope was used for astrometric follow up of twilight survey targets with the Goodman High Throughput Spectrograph in imaging mode (observatory code I33) with a 7.2~arcmin diameter field of view and a pixel scale of 0.15~arcsec/pixel \citep[][]{Clemens2004}. The follow up data were taken in the SDSS r-band filter, and the telescope was non-sidereally tracked at a rate consistent with the target's on-sky motion. I33 was used to follow up (594913) \an on 2020-01-08 and 2020-11-24.

\section{Results}

In total, 1 Aylo, 3 Atiras, and 8 comets were discovered in the twilight survey between 2019 September and 2022 September. A list of the one Aylo, three Atiras, and eight comets and their orbital elements discovered during the ZTF twilight survey is shown in Table~1, and their observational circumstances are listed in Table~2. Example discovery images are shown in in Fig.~9 for 2021 PB$_2$ and P/2022 P2. Cartesian representations of the orbits of the Aylo and Atira discoveries and comet discoveries are shown in Figs.~10 and 11. Additionally, the twilight survey serendipitously recovered 1 Aylo, 12 Atiras, and 6 comets, including interstellar comet 2I/Borisov \citep[][]{Bolin20202I, Bolin2020HST}. The comets were recovered using the Tails neural network pipeline as described in previous sections. The list of Aylo, Atira, and comet recoveries is listed in Table 3, and their observational circumstances are listed in Table 4. In addition, 45,536 unique known Main Belt asteroids and 265 known unique near-Earth objects were recovered in the twilight survey. There were no discoveries of new Amors, Apollos, or Atens due to the bright limiting magnitude of the twilight survey, the relatively completeness of objects in these classes by other surveys, and observational selection effects that favor the detection of these objects at large solar elongation distances \citep[][]{Jedicke2016}.

\begin{figure}\centering
\includegraphics[width=0.75\linewidth]{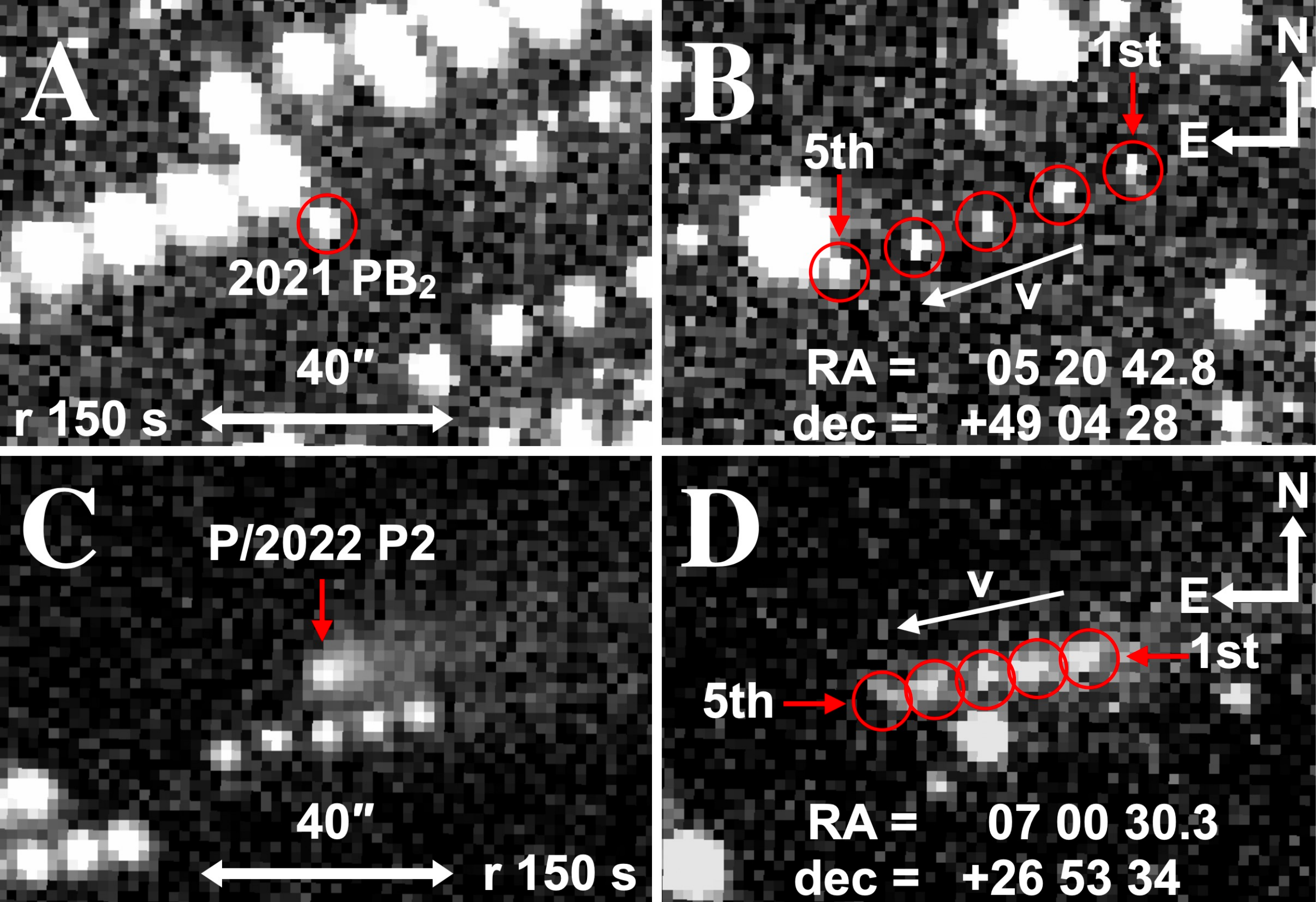} 
\caption{\textbf{P48/ZTF discovery images of 2021 PB$_2$ and P/2022 P2.} \textbf{Panel A:} a composite stack of the detections of 2021 PB$_2$ taken on 2021-08-03 in all five r -band images with an equivalent total exposure time of 150 s. All five exposures were aligned and stacked on the asteroid. \textbf{Panel B:} composite of the discovery images of 2021 PB$_2$ stacked on the background stars showing all five discovery detections. The images were aligned on the background stars
before being coadded. The first and fifth positions of the asteroid are labeled. The asteroid was moving $\sim$2.3 arcsec min$^{-1}$ in the south-east direction. \textbf{Panel C:} a composite stack of the 5 x 30 s r-band discovery images of P/2022 P2 taken on 2022-08-15 where the images were aligned and stacked on the comet. The comet has a $\sim$20 arcsec long tail pointing north of west and a coma $\sim$5 arcsec wide. \textbf{Panel D:} a composite stack of the discovery images of P/2022 P2 where the images were aligned and stacked on the background stars. The first and fifth positions of the comet are labeled. The comet was moving south of east at $\sim$1.4 arcsec min$^{-1}$. The skyplane motion and cardinal directions are indicated on panels B and D, and the image scale is indicated on panels A and C. The detections are encircled in red or indicated by a red circle.}
\end{figure}

\begin{figure}\centering
\includegraphics[width=0.4596\linewidth]{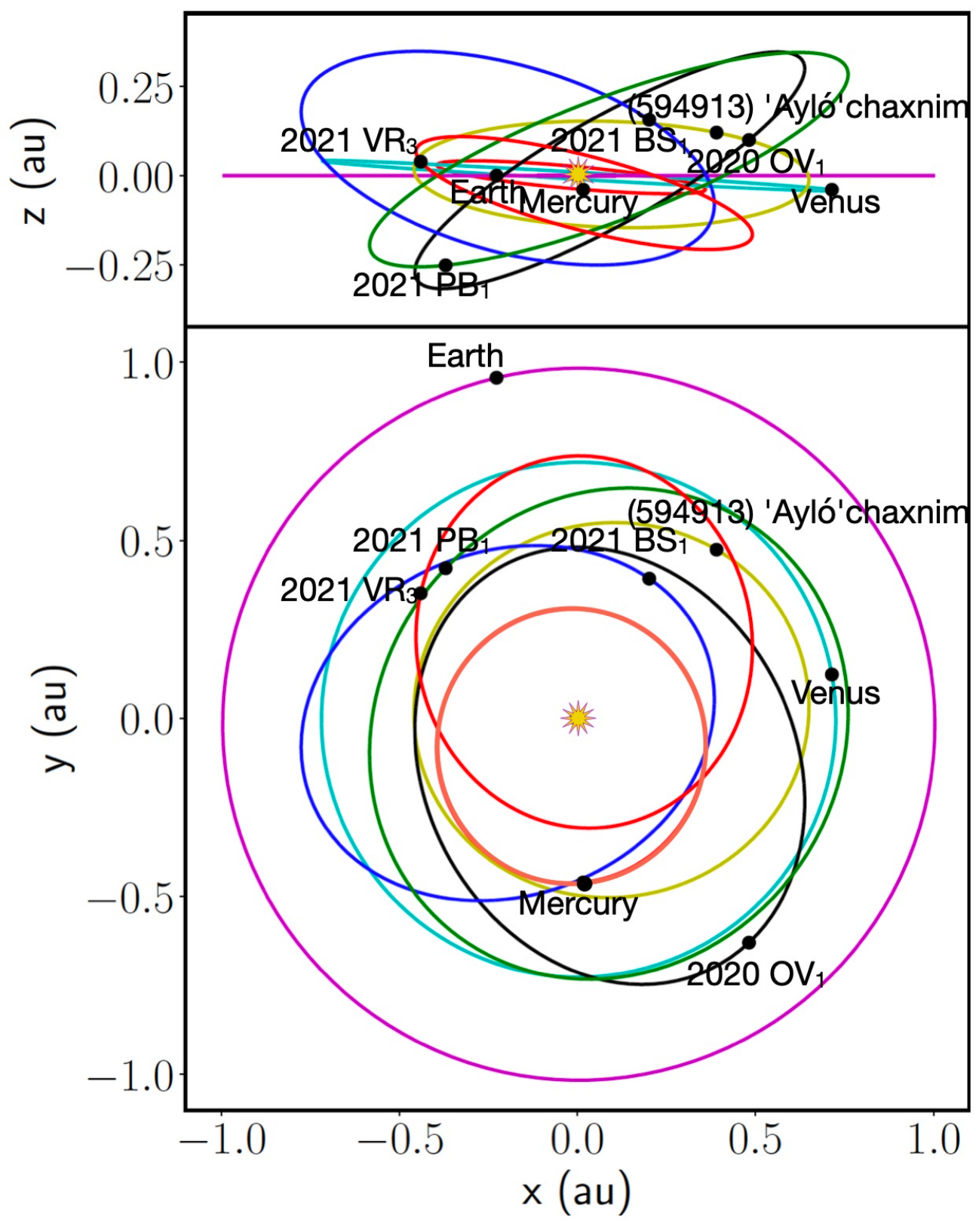}
\caption{\textbf{Orbital diagram of P48/ZTF twilight survey Aylo and Atira discoveries.} \textbf{Top panel:} a side-view snapshot of the solar system on 2020 Jan 04 showing the orbits of Mercury (pink), Venus (aqua), Earth (purple) (594913) \an (yellow), 2020 OV$_1$ (black), 2021 BS$_1$ (blue), 2021 PB$_1$ (green), and 2021 VR$_3$ (red).  \textbf{Bottom panel:} the same as the top panel but looking from above the solar system's orbital plane. The heliocentric Cartesian coordinates x, y, and z are indicated by the position of the Sun as the origin.}
\end{figure}

\begin{figure}\centering
\includegraphics[width=0.45\linewidth]{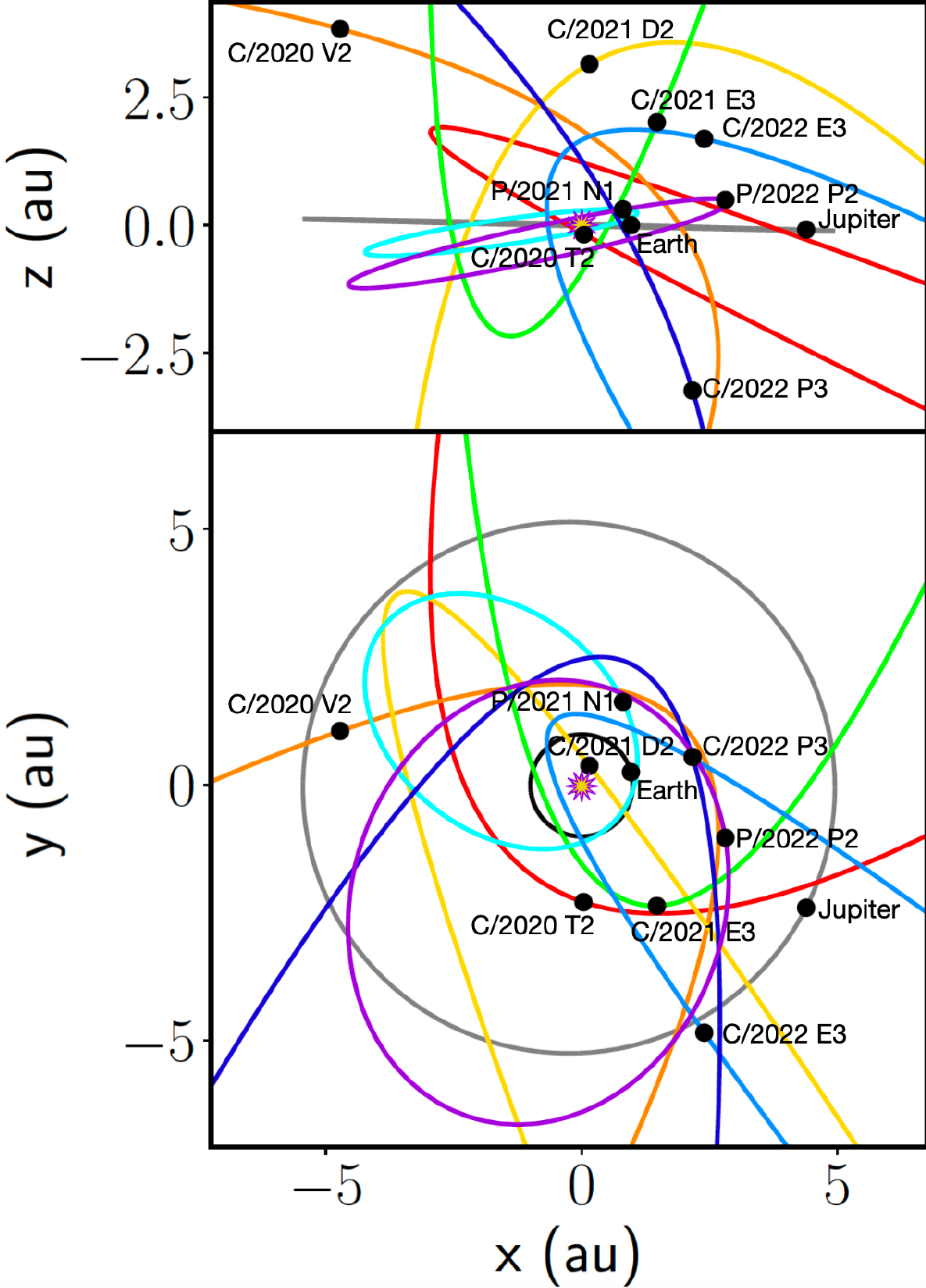}  
\caption{\textbf{Orbital diagram of P48/ZTF twilight survey comet discoveries.} \textbf{Top panel:} a side-view snapshot of the solar system on 2021 Oct 7 showing the orbits of Earth (black), Jupiter (gray), C/2020 T2 (red), C/2020 V2 (orange), C/2021 D2 (yellow), C/2021 E3 (green), P/2021 N1 (cyan), C/2022 E3 (light blue), C/2022 P3 (dark blue), and P/2022 P2 (purple).  \textbf{Bottom panel:} the same as the top panel but looking from above the solar system's orbital plane. The heliocentric Cartesian coordinates x, y, and z are indicated by the position of the Sun as the origin.}
\end{figure}

The Sun-centric ecliptic positions of twilight survey Aylo, Atira, and comet discoveries and recoveries are shown in Fig.~2. Overall, the morning portion of the twilight survey resulted in more discoveries where only (594913) \an was discovered in the evening twilight survey. This may be due to the greater ease of recovering objects in the morning twilight sky since more time is available to recover objects before they go into solar conjunction compared to the evening twilight sky \citep[][]{Bolin20202I}. More recoveries occurred in the evening twilight sky compared to discoveries (7 recoveries vs 1 discovery). However, the morning twilight survey resulted in more discoveries compared to recoveries (12 discoveries vs. 11 recoveries). 

The Sun-centric angular distance and magnitude of the twilight survey Aylo, Atira, and comet discoveries and recoveries are shown in Fig.~3. The smallest Sun-centric angular distance Aylo observation was taken of (594913) \an at 39.5$^{\circ}$ from the Sun during evening twilight on 2020 Jan 4, and the smallest Sun-centric angular distance Atira observation was taken at 32.7$^{\circ}$ from the Sun during evening twilight for the recovery of (413563) on 2021 Feb 07. The smallest Sun-centric angular distance comet observation was made to discover P/2022 P2 on 2022 Aug 15 during morning twilight. As seen in the bottom right panel of Fig.~3, twilight survey asteroid and comet discoveries occurred between an r magnitude of $\sim$17.2 to $\sim$20.7.

Figs.~4-6 show the time after/before sunset, altitude of the Sun, and the Sun-centric angular distance of the twilight survey Aylo, Atira, and comet discoveries and recoveries plotted versus limiting magnitude. In general, fainter objects were found further from the Sun/longer before or after the Sun set or rose. The Aylo, Atiras, and comets were found 0.5-1 magnitudes brighter than the limiting magnitude of the images they were found in. Fig.~7 and 8 show the day of the year versus the Sun-centric angular distance twilight survey when Aylo, Atira, and comet discoveries and recoveries were made. In general, evening twilight survey discoveries and recoveries favor the first half of the year (top left panels of Figs.~7 and 8) whereas Aylo, Atira, and comet discoveries and recoveries are more evening distributed through the year for the morning portion (top right and bottom panels of Figs.~7 and 8).

The NEOMOD NEO model \citep[][]{Nesvorny2023NEOMOD} was used to generate the semi-major axis, a, eccentricity, e, inclination, i, aphelion, Q, and absolute magnitude, H, distributions of the Aylo and Atira populations as shown in Fig.~12. The orbital elements and H of Aylo and Atira discoveries and recoveries made by the twilight survey are overplotted. The range covered by twilight survey Aylo and Atira discoveries and recoveries is between $\sim$0.5 au and $\sim$0.8 au, the e range is between $\sim$0.15 and $\sim$0.5 with the Q ranging between $\sim$0.65 au and 0.95 au. The i range is between 15 and 50$^{\circ}$, missing some of the lower i Atiras in the population owing to the higher ecliptic latitude coverage of the twilight survey \citep[][]{Jedicke2016}. The albedo model  \citep[][]{Morbidelli2020albedo} was used with orbital elements of the Aylo and Atira asteroids observed in the twilight survey to estimate their albedos and are listed in Tables~1 and 3. Note the mismatch in the albedo estimated for (163913) and the albedo inferred by radar diameter measurements \citep[][]{Rivera-Valentin2017}.

\begin{figure}\centering
\includegraphics[width=0.49\linewidth]{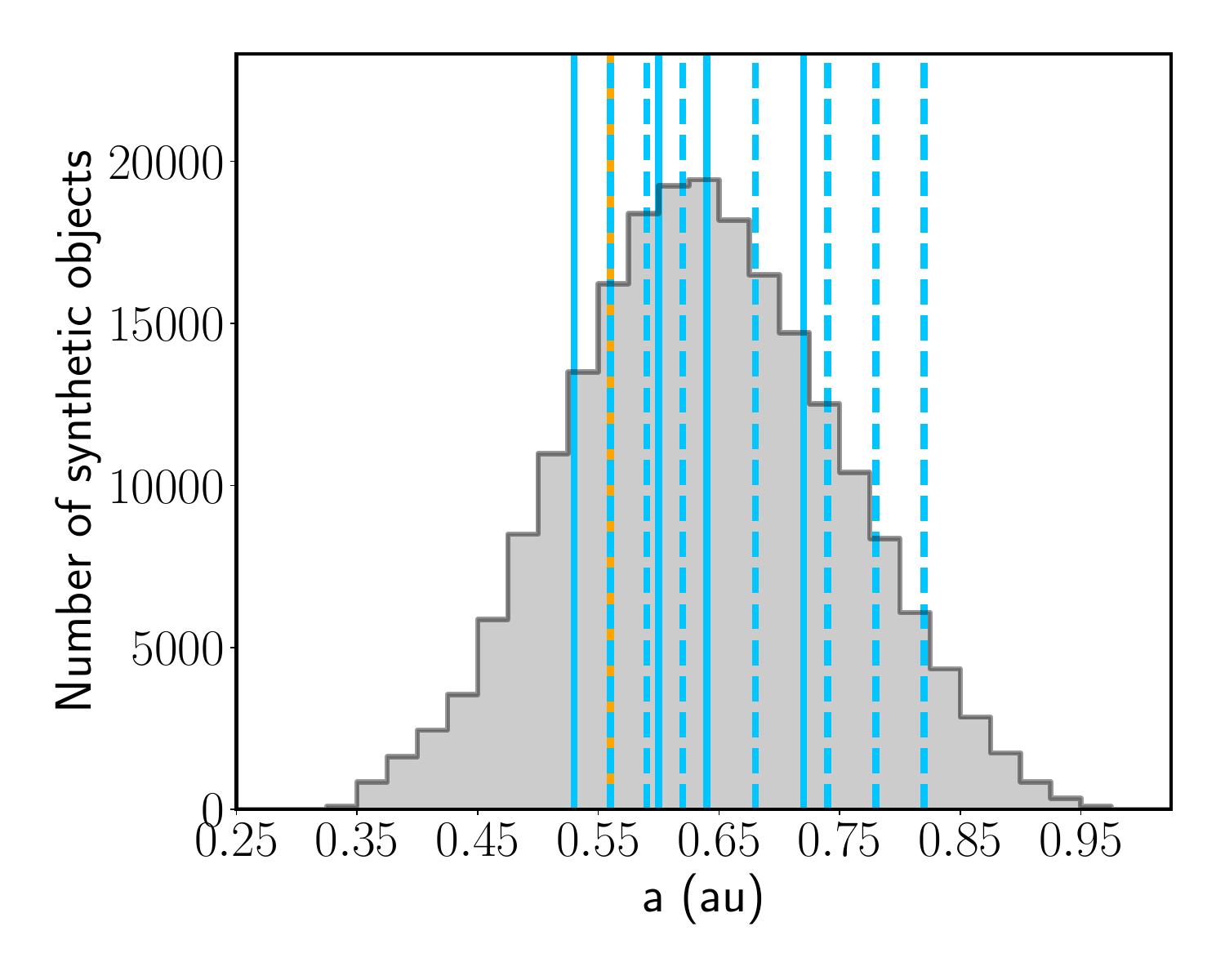} 
\includegraphics[width=0.49\linewidth]{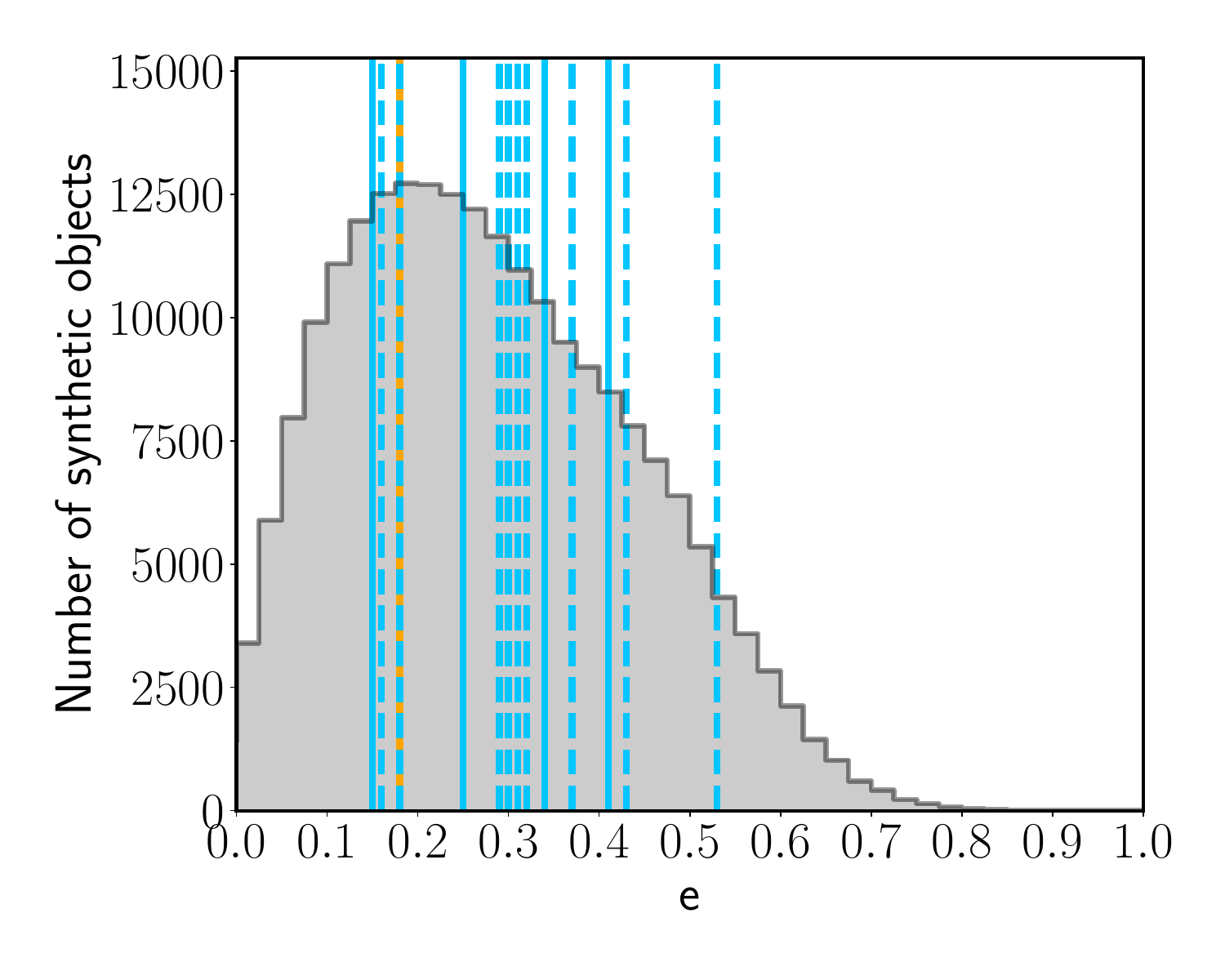}
\includegraphics[width=0.49\linewidth]{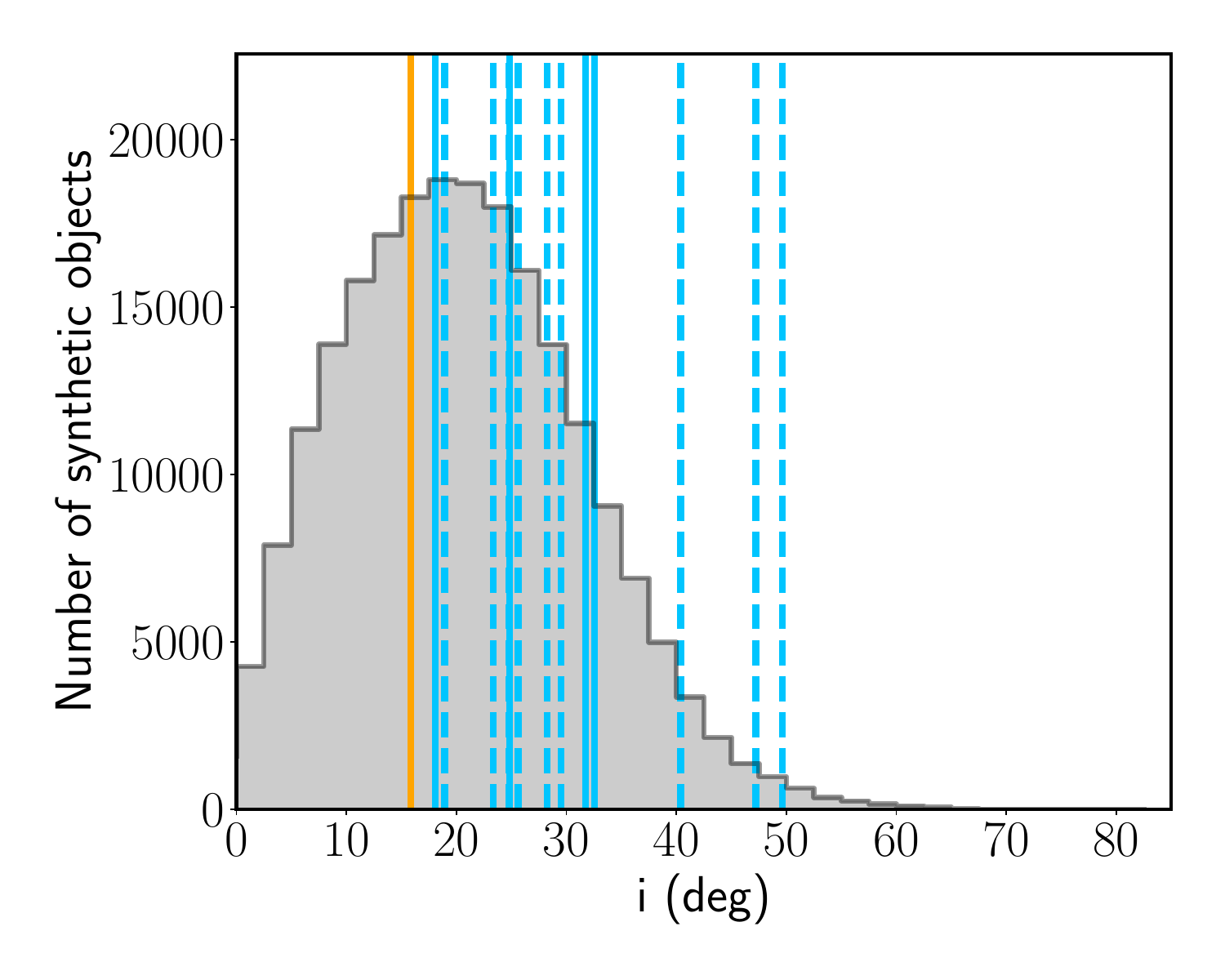} 
\includegraphics[width=0.49\linewidth]{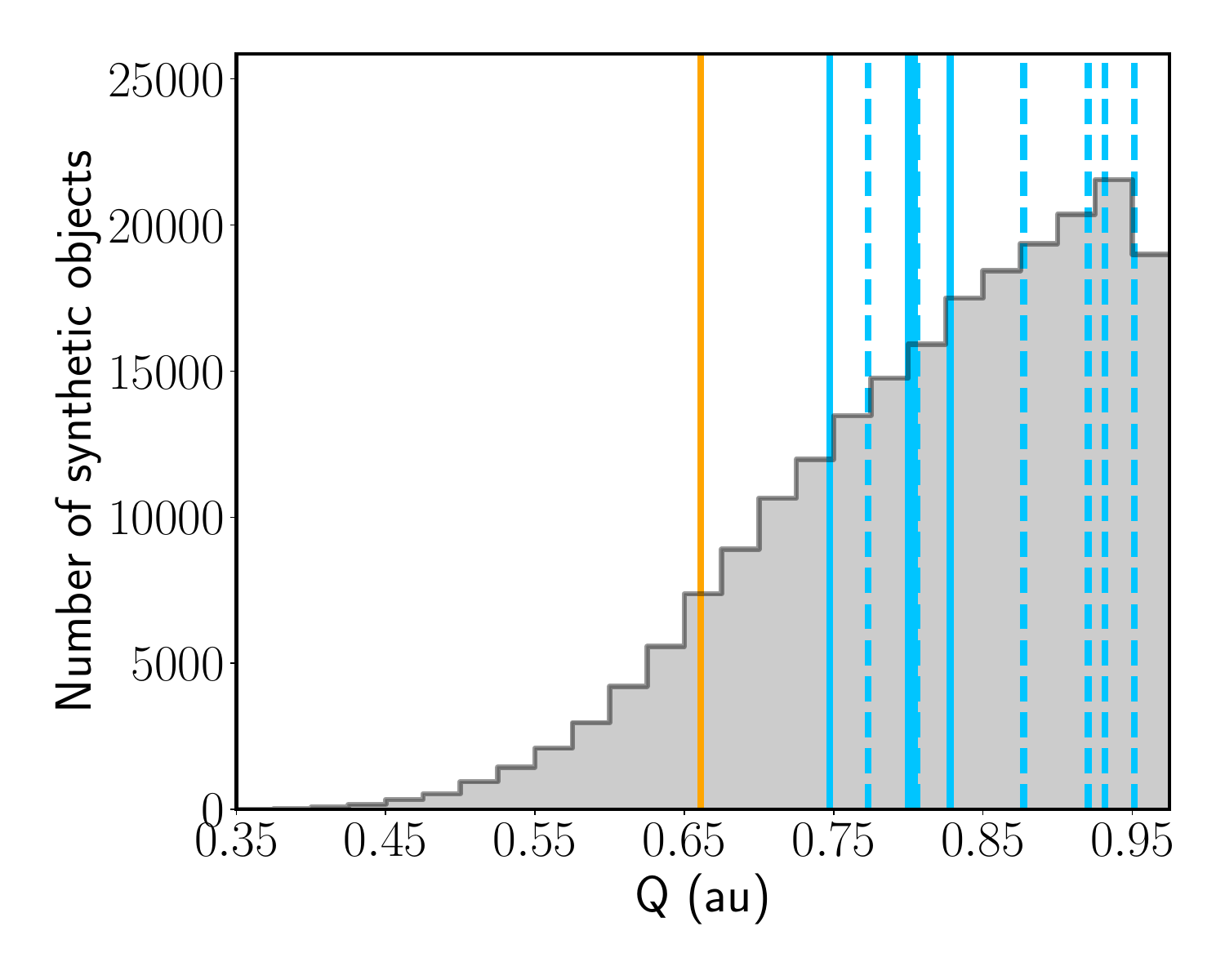}
\includegraphics[width=0.49\linewidth]{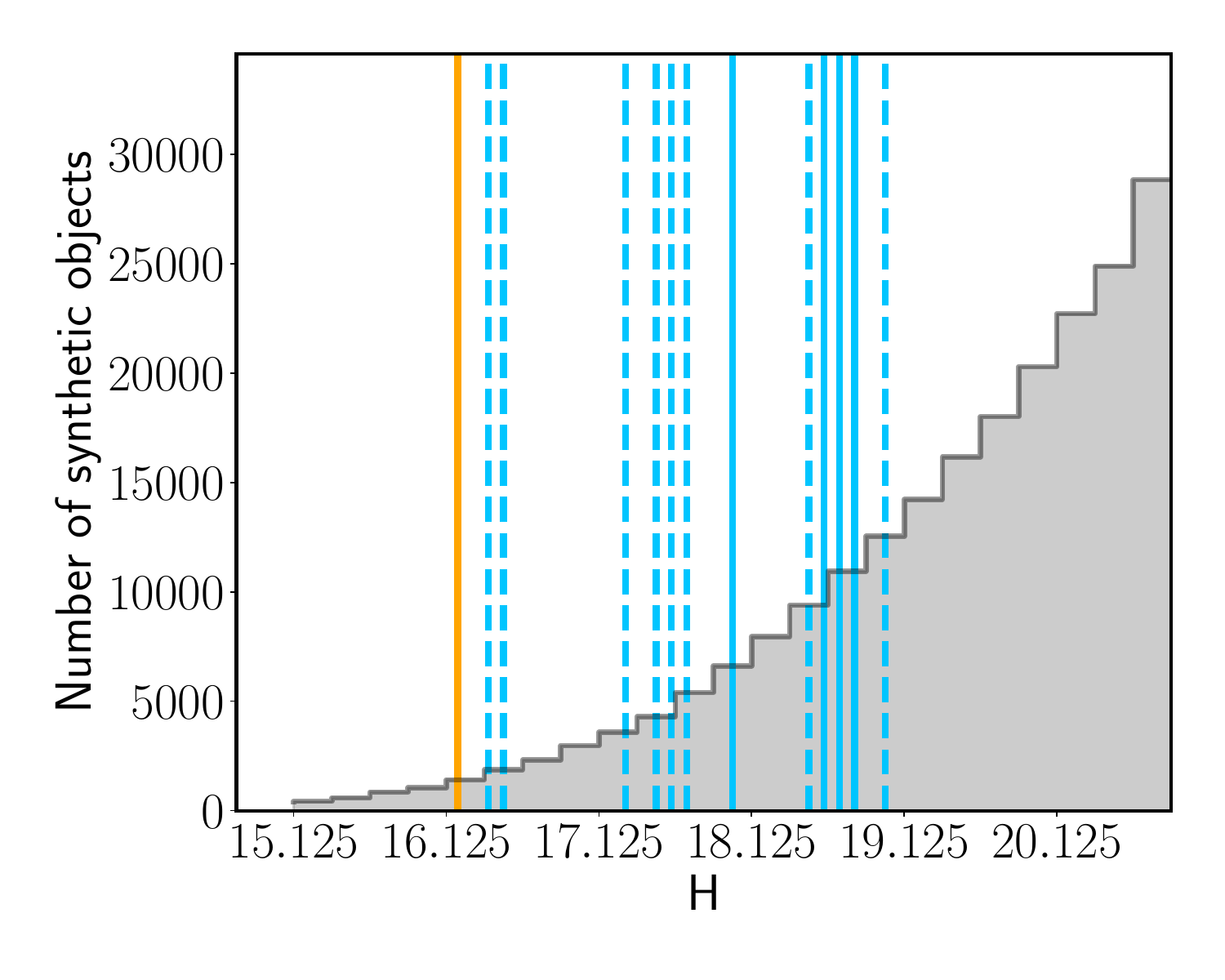} 
\caption{\textbf{a, e, i, Q and H histograms for the Aylo and Atira populations from the NEOMOD population model.} \textbf{Top left panel:} semi-major axis, a, histogram for the Aylo and Atira populations with a bin size of 0.025 au. \textbf{Top right panel:} e histogram for the Aylo and Atira populations with a bin size of 0.025. \textbf{Second row left panel:} i histogram for the Aylo and Atira populations with a bin size of 2.5$^{\circ}$. \textbf{Second-row right panel:} Q histogram for the Aylo and Atira populations with a bin size of 0.025 au. \textbf{Bottom panel:} H histogram for the Aylo and Atira populations with a bin size of 0.25 mag. Orange and blue lines indicate the orbital element and H value of Aylo and Atira discovery and recoveries. The lines will be solid for ZTF discoveries and dashed for recoveries. The sample of synthetic Aylos and Atiras drawn from NEOMOD is 217730. }
\end{figure}

\section{Discussion and conclusion}
The Palomar twilight survey, between 2019 September and 2022 September, demonstrated the capabilities for discovering near-sun asteroids and comets such as Aylos, Atiras, and long and short-period comets. Notable discoveries include the first-known asteroid located entirely within the orbit of Venus, (594913) \an \cite{Bolin2022IVO}, naked-eye bright comet \cite[][]{Bolin2024E3}, and four additional Atira asteroids. The discovery of 5 inner Atira/Aylo objects by ZTF between 2019 and 2022 September represents $\sim$56$\%$ of all Atira/Aylo discoveries in this time period. In this same time period, two Atira objects were found using the V\'{i}ctor M. Blanco 4-meter Telescope at Cerro Tololo Inter-American Observatory \cite{Sheppard2022atira}, and two were found by the Catalina Sky Survey \cite{Zavodny2008,ChristensenCSS2012}.  Additionally, it will serve as a test example of future surveys exploring observations at small Sun-centric angular distances. The future Rubin Observatory Legacy Survey of Space and Time (LSST) \citep[][]{Ivezic2019} will execute a near-Sun twilight survey starting in its first year of observations \cite{PST2023,Jones2023}. 

Like the Palomar twilight survey, the Rubin Observatory twilight survey will focus on observing fields near the Sun during twilight. The top panel of Fig.~13 shows the cumulative distribution of the Sun-centric angular distance for simulated LSST twilight survey fields. Almost all the pointings are within 46$^{\circ}$ of the Sun, inside the maximum Sun-centric angular distance of Aylos, and will visit each field three times \cite[]{Schwamb2023ApJS,Jones2024}. Additionally, over a 10-year time-space, the Rubin twilight survey is expected to find $\sim$50$\%$ of all Aylos with H$<$20 \cite[][]{Jones2023} providing unprecedented coverage of the Aylo population. The average per year skyplane coverage of the LSST and ZTF near-Sun twilight surveys are shown in middle and bottom panels of Fig.~13. The LSST twilight survey covers ecliptic latitudes between -40$^{\circ}$ and 20$^{\circ}$ ecliptic latitude while the ZTF twilight survey covers between -20$^{\circ}$ and 50$^{\circ}$ ecliptic latitude. The average number of pointings per sq. deg. per year covered by the LSST twilight survey is $\sim$3 pointings ranging between 0 and 10 pointings (5th and 95th percentile) with a peak of 14 pointings. The ZTF twilight survey by comparison covers an average number of pointings per sq. deg. per year of $\sim$12 ranging between 0 and 40 pointings with a peak of 62 pointings. While the amount of sky area covered by the LSST twilight survey is significantly smaller than the amount of sky covered by the LSST survey,  LSST will go much deeper compared to ZTF, reaching a limiting magnitude of r$\gtrsim$23 in the twilight survey fields and will be able to detect Aylos with absolute magnitude as faint as H$\sim$22 \cite{Jones2023}. By comparison, (594913) \anns, detected by ZTF, has an H$\sim$16 \cite{Bolin2022IVO}.

\begin{figure}\centering
\includegraphics[width=0.5\linewidth]{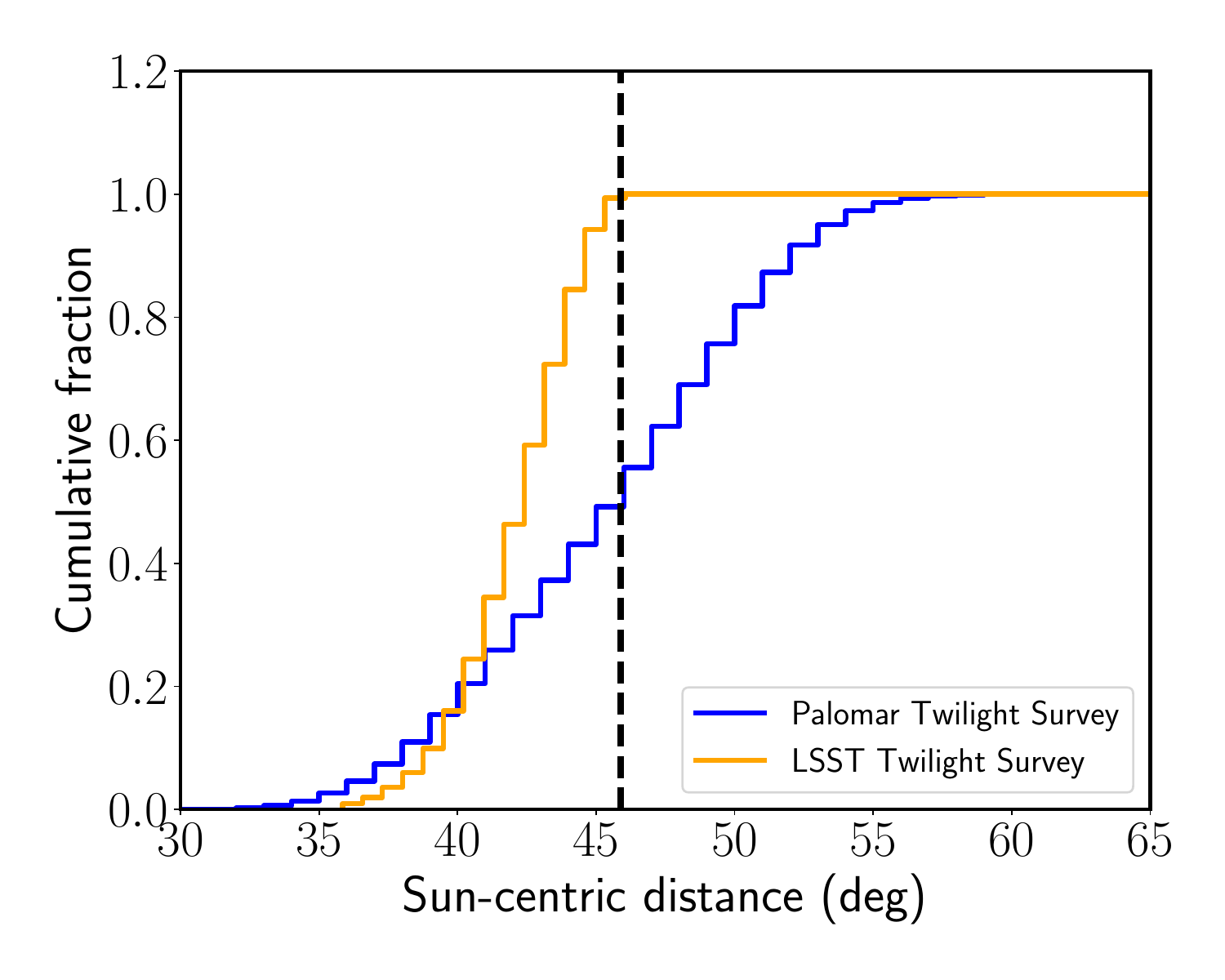}
\includegraphics[width=0.66\linewidth]{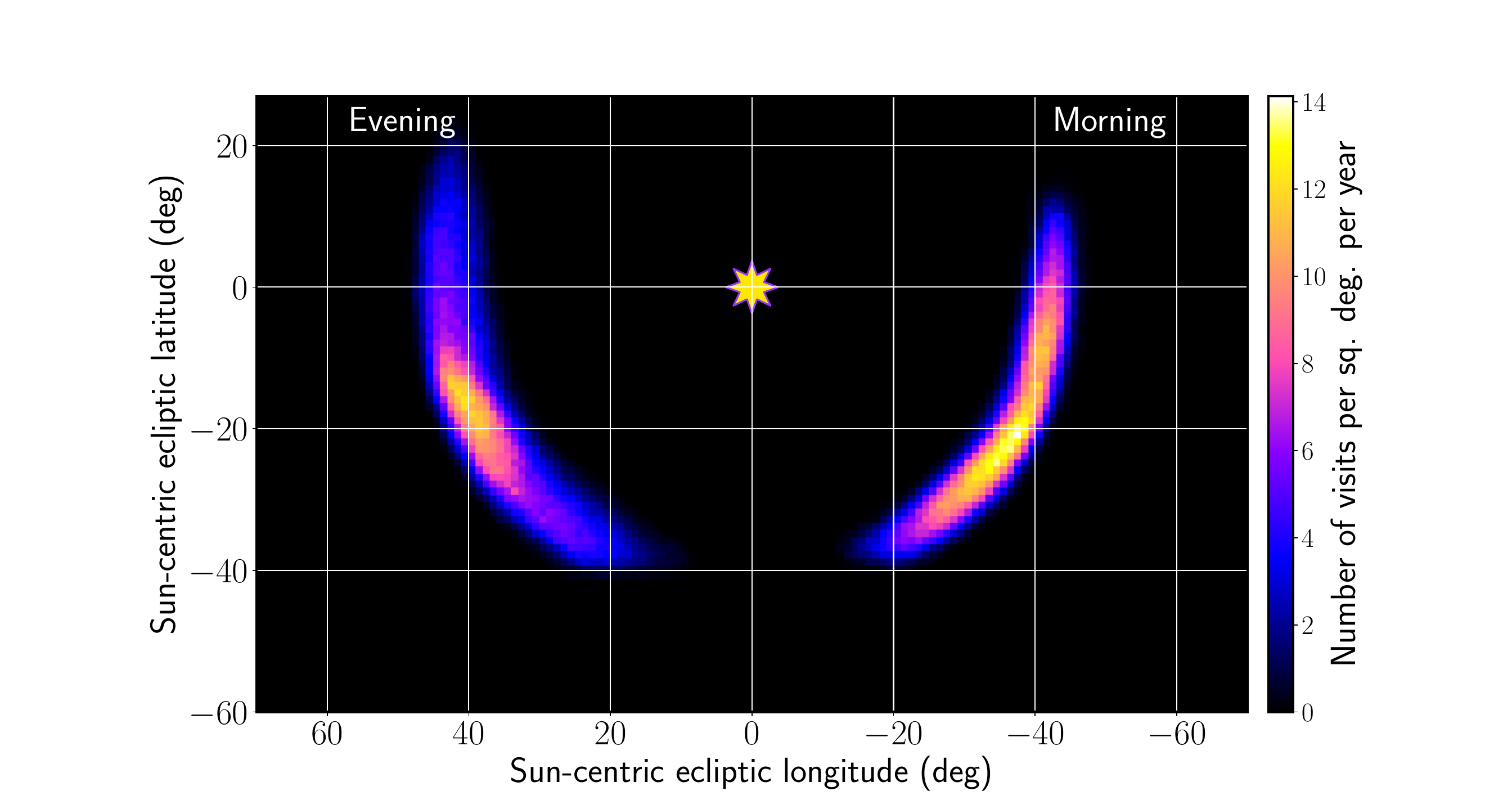}
\includegraphics[width=0.66\linewidth]{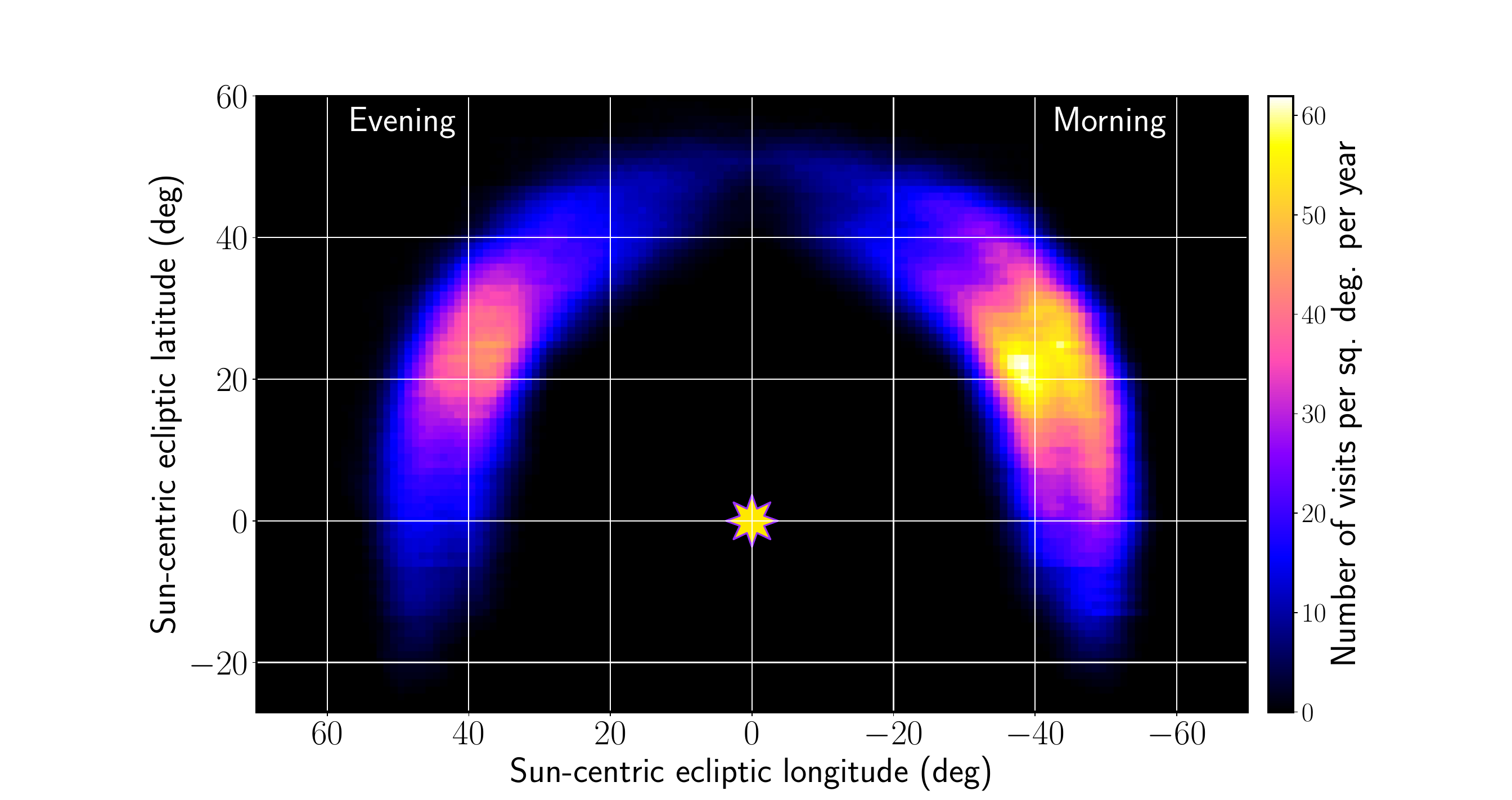}
\caption{\textbf{Cumulative Sun-centric angular distance and limiting magnitude distributions of the Palomar twilight survey and the Cadence Baseline 3.4 LSST near-Sun twilight survey, and the average pear year sky coverage of the LSST and ZTF near-Sun twilight surveys.} \textbf{Top panel:} The cumulative Sun-centric angular distance distribution of evening and morning Palomar twilight survey fields taken between 2019 September and 2022 March for the evening portion and 2019 September and 2022 September for the morning portion are plotted in blue. The 10-year Baseline 3.4 cadence is plotted in orange, adapted from \cite[][]{Jones2023,Jones2024}. The maximum Sun-centric angular distance of Aylo asteroids is plotted as a vertical black dashed line. \textbf{Middle panel:} the average per year Sun-centric ecliptic sky-plane distribution of the 10 y Cadence Baseline 3.4 LSST near-Sun twilight survey fields \cite[][]{Jones2024}. \textbf{Bottom panel:} the average per year Sun-centric ecliptic sky-plane distribution of the ZTF twilight survey fields between 2019 September-2022 September. The color bars in the bottom two panels indicate the number of visits per sq.deg. per year covered by the ZTF and LSST twilight surveys. The position of the Sun is indicated by a yellow starburst.}
\end{figure}

In addition to hunting for near-Sun asteroids and comets, twilight surveys have a variety of applications in astrophysics. By varying the exposure time during twilight to exposures as short as 0.1 seconds, the dynamic range of the Rubin Observatory survey can be increased, as low 11th magnitude increasing the dynamic range to over 6 magnitudes of flux \cite[][]{Stubbs2013}. This will have the added benefit of being sensitive to detecting bright transients such as supernovae and bright stars in the galaxy \citep[][]{Fremling2020, Perley2020, Burdge2020, Rehemtulla2024}. Additionally, the coverage of the twilight survey could be expanded by taking fewer exposures per footprint. This would fall short of the minimum of three exposures needed to measure angular velocity differences between pairs of asteroid detections \cite[][]{Milani2010}. However, the source of a third exposure could be made with an auxiliary telescope following up on candidates in real-time. Thus, future near-Sun twilight surveys, such as for the upcoming LSST, have the potential to benefit multiple areas of astrophysics and planetary science.

\bibliographystyle{Science}
\bibliography{/Users/bolin/Dropbox/Projects/NEOZTF_NEOs/neobib}

\section*{Acknowledgements}

\noindent The authors thank the anonymous reviewers, and the editor, Lauri Faega, for their comments and guidance in improving the manuscript. Additionally, the authors thank R. S. Walters for their help in follow up of comets and asteroids form Palomar Observatory and C.W. Stubbs for helpful discussion on the Rubin twilight survey. The authors wish to recognize and acknowledge the cultural significance that Palomar Mountain has for the Pauma Band of the Luise\~{n}o Indians. The data presented in this paper are available in ZTF data release 19. Based on observations obtained with the Samuel Oschin Telescope 48-inch and the 60-inch Telescope at the Palomar Observatory as part of the Zwicky Transient Facility project. ZTF is supported by the National Science Foundation under Grants No. AST-1440341 and AST-2034437 and a collaboration including current partners Caltech, IPAC, the Weizmann Institute for Science, the Oskar Klein Center at Stockholm University, the University of Maryland, Deutsches Elektronen-Synchrotron and Humboldt University, the TANGO Consortium of Taiwan, the University of Wisconsin at Milwaukee, Trinity College Dublin, Lawrence Livermore National Laboratories, IN2P3, University of Warwick, Ruhr University Bochum, Northwestern University and former partners the University of Washington, Los Alamos National Laboratories, and Lawrence Berkeley National Laboratories. Operations are conducted by COO, IPAC, and UW. C.A. 

\noindent C.F.~acknowledges support from the Heising-Simons Foundation (grant $\#$2018-0907). 

\noindent M.~W.~Coughlin acknowledges support from the National Science Foundation with grant numbers PHY-2308862 and PHY-2117997.

\noindent Based in part on observations obtained at the Southern Astrophysical Research (SOAR) telescope, which is a joint project of the Minist\'{e}rio da Ci\^{e}ncia, Tecnologia e Inova\c{c}\~{o}es (MCTI/LNA) do Brasil, the US National Science Foundation's NOIRLab, the University of North Carolina at Chapel Hill (UNC), and Michigan State University (MSU).

\noindent CMC receives funding from UKRI grant numbers ST/X005933/1 and ST/W001934/1. The Liverpool Telescope is operated on the island of La Palma by Liverpool John Moores University in the Spanish Observatorio del Roque de los Muchachos of the Instituto de Astrofisica de Canarias.

\noindent Part of this work was carried out at the Jet Propulsion Laboratory, California Institute of Technology, under contract with NASA 80NM0018D0004.

\clearpage
\newpage
\begin{landscape}
\begin{table}[]
\caption{Aylo and Atira asteroids, and Long Period and Short Period comets discovered by the P48. Orbital elements provided by the JPL Small-Body Database Lookup accessed on 2023 September 30.}
\centering
\begin{tabular}{llllllllllllll}
\hline
object      & date$^{1}$ & class$^{2}$ & a/q$^{3}$    & P$^{4}$ & e$^{5}$    & i$^{6}$         & Q$^{7}$      & H,M1$^{8}$  & p$_v^{9}$ & D$^{10}$ & arc$^{11}$  & AI$^{12}$  & ref. \\
            &      (UTC)          &       & (au)   & (y)    &      & ($^{\circ}$) & (au)   & (mag) &        & (m)      & (d)  &     &           \\ \hline
(594913) \an & 2020-01-04     & Aylo    & 0.56   & 0.41   & 0.18 & 15.87     & 0.65   & 16.2  & 0.21   & 1670     & 1351 & N/A &   \cite[][]{Bolin2022IVO}        \\
2020 OV$_1$ & 2020-07-19     & Atira & 0.64   & 0.51   & 0.25 & 32.58     & 0.80   & 18.7  & 0.21   & 530      & 1179 & N/A &     \cite[][]{Bolin2020MPECOV1}      \\
C/2020 T2 (Palomar)   & 2020-10-07     & Long  & 2.05 & 5529.0 & 0.99 & 27.87     & 623.30 & 9.4   & N/A    & N/A      & 1015 & Yes &     \cite[][]{Duev2020MPECT2}      \\
C/2020 V2 (ZTF)  & 2020-11-02     & Long  & 2.23   & undef. & 1.0  & 131.61    & undef. & 8.5   & N/A    & N/A      & 1241 & Yes &     \cite[][]{Bolin2020MPECV2}      \\
2021 BS$_1$ & 2021-01-14     & Atira & 0.60   & 0.46   & 0.34 & 31.73     & 0.80   & 18.6  & 0.21   & 550      & 859  & N/A &     \cite[][]{Bolin2021MPECBS1}      \\
C/2021 D2 (ZTF)  & 2021-02-19     & Hyperbolic  & 2.94   & undef. & 1.0  & 83.83     & undef. & 13.3  & N/A    & N/A      & 792  & Yes &    \cite[][]{Bolin2021MPECD2}       \\
C/2021 E3 (ZTF)  & 2021-03-09     & Long  & 1.78   & undef. & 1.0  & 112.56    & undef. & 11.1  & N/A    & N/A      & 780  & Yes &     \cite[][]{Bolin2021MPECE3}      \\
P/2021 N1 (ZTF)  & 2021-07-02     & Short & 0.96   & 5.14   & 0.68 & 11.51     & 4.99   & 17.8  & N/A    & N/A      & 25   & Yes &     \cite[][]{Bolin2021MPECN1}      \\
2021 PB$_2$ & 2021-08-03     & Atira & 0.72   & 0.61   & 0.15 & 24.83     & 0.83   & 18.8  & 0.21   & 510      & 3392 & N/A &       \cite[][]{Bolin2021MPECPB2}    \\
2021 VR$_3$ & 2021-11-03     & Atira & 0.53   & 0.39   & 0.41 & 18.06     & 0.75   & 18.0  & 0.21   & 720      & 1012 & N/A &      \cite[][]{Bolin2021MPECVR3}     \\
C/2022 E3 (ZTF)  & 2022-03-02     & Long  & 1.11   & undef. & 1.0  & 109.17    & undef. & 10.8  & N/A    & N/A      & 672  & Yes &    \cite[][]{Bolin2024E3}       \\
C/2022 P3 (ZTF)  & 2022-08-02     & Long  & 2.56   & 3794.2 & 0.99 & 59.52     & 483.96 & 14.5  & N/A    & N/A      & 261  & Yes &    \cite[][]{Bolin2022MPECP3}       \\
P/2022 P2 (ZTF)  & 2022-08-15     & Short & 4.48   & 9.5    & 0.56 & 12.44     & 6.99   & 9.2   & N/A    & N/A      & 230  & Yes & \cite[][]{Bolin2022MPECP2}
\end{tabular}
\begin{tablenotes}
\item \textbf{Notes.} (1) Date of discovery, (2) dynamical class, (3) semi-major axis (for asteroids)/perihelion (for comets), (4) orbital period, (5) eccentricity, (6) inclination, (7) aphelion, (8) absolute magnitude (for asteroids)/comet total magnitude (for comets),  (9) visible albedo inferred from albedo model \citep[][]{Morbidelli2020albedo}, (10) diameter inferred from albedo and absolute magnitude, (11) observational arc for orbit in JPL HORIZONS, (12) flag for whether comet was identified with the Tails pipeline.
\end{tablenotes}
\end{table}
\end{landscape}

\clearpage
\newpage
\begin{landscape}
\begin{table}[]
\caption{Observational circumstances of Aylo and Atira asteroids, and Long Period and Short Period comets discovered by the P48.}
\centering
\begin{tabular}{lllllllllllll}
\hline
object      & date$^{1}$ & time$^{2}$ & $E$$^{3}$ & $\lambda_{\odot}$$^{4}$ & $\beta_{\odot}$$^{5}$ & $\alpha$$^{6}$     & R$^{7}$    & $\Delta$$^{8}$ & $X$$^{9}$ & $\delta_{\theta}$$^{10}$   & $m_{lim}$$^{11}$ & m$_r$$^{12}$ \\
            &      (UTC)          &                 & ($^{\circ}$)    & ($^{\circ}$)           & ($^{\circ}$)          & ($^{\circ}$) & (au) & (au)  &         & (\SI{}{ arcsec}) & (mag)        & (mag) \\ \hline
(594913) \an & 2020-01-04     & evening         & 39.5         & 37.0                & 8.6                & 86.4      & 0.6 & 0.8  & 2.3     & 1.7     & 19.2         & 18.0  \\
2020 OV$_1$ & 2020-07-19     & morning         & 48.2         & -44.7               & 24.4               & 106.2     & 0.8 & 0.5  & 2.0    & 1.8    & 20.3         & 20.1  \\
C/2020 T2 (Palomar)   & 2020-10-07     & morning         & 43.3         & -39.0               & 21.2               & 10.3      & 3.7 & 4.4  & 2.5   & 2.0    & 20.6        & 19.3  \\
C/2020 V2 (ZTF)   & 2020-11-02     & morning         & 44.2         & -37.2               & 28.4               & 43.9      & 8.7 & 9.4  & 2.5   & 2.1    & 20.2        & 19.0  \\
2021 BS$_1$ & 2021-01-14     & morning         & 49.3         & -43.1               & 29.6               & 89.5      & 0.7 & 0.7  & 2.0   & 2.3    & 20.3        & 19.5  \\
C/2021 D2 (ZTF)   & 2021-02-19     & morning         & 45.9         & -14.0               & 44.6               & 9.11      & 4.5 & 5.1  & 2.2   & 2.1    & 20.0        & 19.2  \\
C/2021 E3 (ZTF)   & 2021-03-09     & morning         & 50.8         & -24.8               & 46.6               & 8.3       & 5.3  & 5.9   & 1.9   & 2.1    & 20.6        & 19.6  \\
P/2021 N1 (ZTF)  & 2021-07-02     & morning         & 53.0         & -54.4               & 7.7                & 52.2      & 1.0 & 1.2  & 2.2   & 1.6    & 20.8        & 18.9  \\
2021 PB$_2$ & 2021-08-03     & morning         & 53.1         & -49.9               & 25.9               & 82.1      & 0.8 & 0.7  & 1.8   & 1.6    & 21.0        & 20.4  \\
2021 VR$_3$ & 2021-11-03     & morning         & 49.4         & -49.5               & -14.6              & 87.1      & 0.8 & 0.7  & 2.3   & 3.1    & 20.2        & 19.3  \\
C/2022 E3 (ZTF)  & 2022-03-02     & morning         & 44.1         & -41.0               & 22.0               & 9.2       & 4.3 & 4.9  & 2.2   & 2.0    & 20.1        & 17.3  \\
C/2022 P3 (ZTF)  & 2022-08-02     & morning         & 41.4         & -43.0               & 1.5                & 15.1      & 2.6 & 0.4  & 2.6   & 2.1    & 20.1        & 19.4  \\
P/2022 P2 (ZTF)  & 2022-08-15     & morning         & 39.0         & -40.5               & 4.2                & 18.4      & 2.0  & 2.7  & 2.4   & 1.7    & 19.8        & 18.5 
\end{tabular}
\begin{tablenotes}
\item \textbf{Notes.} (1) discovery date, (2) time of twilight, (3) Sun-centric angular distance, (4) Sun-centric ecliptic longitude, (5) Sun-centric ecliptic latitude, (6) phase angle, (7) heliocentric distance, (8) geocentric distance, (9) airmass, (10) seeing, (11) limiting magnitude of discovery observations in r band, (12) PSF-fitted r magnitude.
\end{tablenotes}
\end{table}
\end{landscape}

\clearpage
\newpage
\begin{landscape}
\begin{table}[]
\caption{Aylo and Atira asteroids, and Long Period and Short Period comets serendipitously  recovered by the P48. Orbital elements provided by the JPL Small-Body Database Lookup.}
\centering
\begin{tabular}{llllllllllllll}
\hline
object      & date$^{1}$ & class$^{2}$ & a/q$^{3}$    & P$^{4}$ & e$^{5}$    & i$^{6}$         & Q$^{7}$      & H,M1$^{8}$  & p$_v^{9}$ & D$^{10}$ & arc$^{11}$  & AI$^{12}$  & ref. \\
            &      (UTC)          &       & (au)   & (y)    &      & ($^{\circ}$) & (au)   & (mag) &        & (m)      & (d)  &     &           \\ \hline
2I/Borsiov          & 2019-10-15 & interstellar & 2.00 & undef.   & 3.36 & 44.05     & undef.   & 13.7  & N/A       & N/A      & 311 & Yes &   \citep[][]{Bolin20192I}        \\
2020 HA$_{10}$          & 2020-02-24 to 2020-04-02 & Atira & 0.82 & 0.74   & 0.16 & 49.65     & 0.95   & 19.0  & 0.20       & 470      & 4043 & N/A &     \cite[][]{Williams2019MPS1518512}      \\
(418265) 2008 EA$_{32}$ & 2020-02-27 to 2022-04-10 & Atira & 0.62 & 0.48   & 0.30 & 28.26     & 0.80   & 16.5  & 0.21       & 1470     & 5470 & N/A &      \cite[][]{Williams2020MPS1163484}     \\
(163693) Atira     & 2020-04-15 to 2022-07-07 & Atira & 0.74 & 0.64   & 0.32 & 25.62     & 0.98   & 16.4  & 0.20/0.02* & 4800$^{\dagger}$     & 7086 & N/A &    \cite[][]{Williams2020MPS1174476}       \\
2019 LF$_6$           & 2020-04-28 to 2022-04-24 & Atira & 0.56 & 0.41   & 0.43 & 29.5      & 0.79   & 17.3  & 0.21       & 1010     & 1108 & N/A &     \cite[][]{Williams2020MPS1186327}      \\
(613676) 2006 WE$_{4}$  & 2020-07-25               & Atira & 0.78 & 0.70   & 0.18 & 24.77     & 0.93   & 18.8  & 0.21       & 510      & 5614 & N/A &      \cite[][]{Williams2020MPS1210088}     \\
2019 AQ$_3$           & 2020-09-21 to 2021-09-26 & Atira & 0.59 & 0.45   & 0.31 & 47.22     & 0.77   & 17.5  & 0.21       & 910      & 2669 & N/A &    \cite[][]{Williams2020MPS1248822}       \\
414P/STEREO     & 2021-01-04 to 2021-01-05 & Short & 0.53 & 4.67   & 0.81 & 23.38     & 5.06   & 16.0  & N/A        & N/A      & 1779 & Yes &      \cite[][]{Bolin2021414P}     \\
P/2022 BV9 (Lemmon)        & 2021-02-06 to 2022-04-19 & Short & 3.33 & 9.10   & 0.23 & 11.93     & 5.38   & 14.4  & N/A        & N/A      & 795  & No  &    \cite[][]{Williams2020MPC163307}       \\
(413563) 2005 TG$_{45}$ & 2021-02-07 to 2022-06-27 & Atira & 0.68 & 0.56   & 0.37 & 23.33     & 0.94   & 17.6  & 0.21       & 880      & 6189 & N/A &    \cite[][]{Williams2021MPS1371067}       \\
2021 BS$_1$           & 2021-02-07 to 2021-02-08 & Atira & 0.60 & 0.46   & 0.34 & 31.73     & 0.80   & 18.6  & 0.21       & 550      & 859  & N/A &     \cite[][]{Williams2021MPS1446889}      \\
2018 JB$_3$           & 2021-02-07 to 2022-04-15 & Atira & 0.68 & 0.56   & 0.29 & 40.39     & 0.88   & 17.7  & 0.21       & 840      & 2895 & N/A &   \cite[][]{Williams2022MPS1582060}        \\
C/2021 D2 (ZTF)         & 2021-03-09               & Hyperbolic  & 2.94 & undef. & 1.0  & 83.83     & undef. & 13.3  & N/A        & N/A      & 792  & Yes &      \cite[][]{Williams2021MPC134013}     \\
(434326) 2004 JG$_6$ & 2021-05-29               & Atira & 0.64 & 0.51   & 0.53 & 18.94     & 0.97   & 18.5  & 0.21       & 580      & 6227 & N/A &    \cite[][]{Williams2021MPC1469946}       \\
P/2021 N1 (ZTF)       & 2021-07-11             & Short & 0.96 & 5.14   & 0.68 & 11.51     & 4.99   & 17.8  & N/A        & N/A      & 25   & Yes &      \cite[][]{Williams2021MPC158560}     \\
(594913) \an       & 2021-08-04 to 2022-02-14 & Aylo    & 0.56 & 0.41   & 0.18 & 15.87     & 0.65   & 16.2  & 0.21       & 1670     & 1351 & N/A &    \cite[][]{Williams2021MPC1442126}      \\
2021 PB$_2$           & 2021-09-13               & Atira & 0.72 & 0.61   & 0.15 & 24.83     & 0.61   & 18.8  & 0.21       & 510      & 3392 & N/A &     \cite[][]{Williams2021MPC1525599}      \\
C/2021 QM45 (PANSTARRS)       & 2022-02-14 to 2022-08-06 & Long  & 2.77 & 6660.8 & 0.99 & 22.82     & 705.24 & 9.1   & N/A        & N/A      & 591  & Yes &     \cite[][]{Williams2022MPS141248}      \\
\end{tabular}
\begin{tablenotes}
\item \textbf{Notes.} (1) Recover date range, (2) dynamical class, (3) semi-major axis (for asteroids)/perihelion (for comets), (4) orbital period, (5) eccentricity, (6) inclination, (7) aphelion, (8) absolute magnitude (for asteroids)/comet total magnitude (for comets),  (9) visible albedo inferred from albedo model \citep[][]{Morbidelli2020albedo}, (10) diameter inferred from albedo and absolute magnitude, (11) observational arc for orbit in JPL HORIZONS, (12) flag for whether comet was identified with the Tails pipeline. (*) albedo for (163693) Atira derived by combining the diameter measured in radar observations and its absolute magnitude. ($\dagger$) diameter for (163693) Atira measured from radar observations \citep[][]{Rivera-Valentin2017}.
\end{tablenotes}
\end{table}
\end{landscape}

\clearpage
\newpage
\begin{landscape}
\begin{table}[]
\caption{Observational circumstances of Aylo and Atira asteroids, and Long Period and Short Period comets serendipitously recovered by the P48.}
\centering
\begin{tabular}{lllllllllllll}
\hline
object      & date$^{1}$ & time$^{2}$ & $E$$^{3}$ & $\lambda_{\odot}$$^{4}$ & $\beta_{\odot}$$^{5}$ & $\alpha$$^{6}$     & R$^{7}$    & $\Delta$$^{8}$ & $X$$^{9}$ & $\delta_{\theta}$$^{10}$   & $m_{lim}$$^{11}$ & m$_r$$^{12}$ \\
            &      (UTC)          &                 & ($^{\circ}$)    & ($^{\circ}$)           & ($^{\circ}$)          & ($^{\circ}$) & (au) & (au)  &         & (\SI{}{ arcsecond}) & (mag)        & (mag) \\ \hline
2I/Borisov          & 2019-10-15 & morning         & 56.7         & 56.5               & 4.1               & 20.9      & 2.3 & 2.7  & 1.8 & 3.9 & 17.4  & 19.2  \\
2020 HA$_{10}$          & 2020-02-24 to 2020-04-02 & morning         & 46.3         & -8.3                & 45.9               & 88.5      & 0.7 & 0.7  & 1.9-2.4 & 2.3-4.2 & 20.1-20.3  & 20.1-20.2  \\
(418265) 2008 EA$_{32}$ & 2020-02-27 to 2022-04-10 & evening         & 44.1         & 42.0                & -6.5               & 61.9      & 0.8 & 1.1  & 2.2       & 1.5-4.5 & 18.9-20.4  & 17.9-18.1  \\
(163693) Atira    & 2020-04-15 to 2022-07-07 & evening         & 50.8         & 48.0                & 12.2               & 60.3      & 0.9 & 1.1  & 1.8-2.3 & 1.8-2.7 & 19.9-20.8  & 17.5-18.3  \\
2019 LF$_6$           & 2020-04-28 to 2022-04-24 & morning         & 40.2         & -33.9               & 25.4               & 93.2      & 0.7 & 0.7  & 2.2-2.5 & 1.9-2.3 & 20.1-20.6  & 18.8-19.2  \\
(613676) 2006 WE$_{4}$  & 2020-07-25               & morning         & 45.6         & -45.0               & 16.1               & 68.3      & 0.78 & 1.0   & 2.2       & 1.6       & 21.0        & 20.7       \\
2019 AQ$_3$           & 2020-09-21 to 2021-09-26 & morning         & 49.0         & -47.2               & -20.8              & 81.8      & 0.8 & 0.8  & 2.2-2.6 & 1.8-2.1 & 20.0-20.9  & 18.8-19.1  \\
414P/STEREO     & 2021-01-04 to 2021-01-05 & evening         & 43.7         & 41.1                & 9.8                & 81.0      & 0.7 & 0.8  & $\sim$2.1   & 2.4-2.9 & 19.7-20.2  & 18.6-19.1  \\
P/2022 BV9 (Lemmon)         & 2021-02-06 to 2022-04-19 & evening         & 43.2         & 41.2                & 5.7                & 11.6      & 3.4 & 4.0  & 1.7-2.0 & 2.0-2.2 & 19.7-20.7  & 19.6-19.7  \\
(413563) 2005 TG$_{45}$ & 2021-02-07 to 2022-06-27 & evening         & 32.7         & 30.0                & 8.0                & 64.3      & 0.6 & 1.1  & 2.6-2.7 & 2.2-2.8 & 19.5-20.0  & 18.6-19.8 \\
2021 BS$_1$           & 2021-02-07 to 2021-02-08 & morning         & 53.5         & -52.1               & 21.2               & 84.4      & 0.8 & 0.7  & 1.9-2.0 & 1.8-2.1 & 20.6-20.9  & 19.9-20.0  \\
2018 JB$_3$           & 2021-02-07 to 2022-04-15 & morning         & 38.7         & -32.7               & 24.4               & 66.2      & 0.7 & 1.0  & 1.8-2.4 & 1.7-2.1 & 19.8-20.2  & 18.8-19.2  \\
C/2021 D2 (ZTF)          & 2021-03-09               & morning         & 51.5         & -25.9               & 46.9               & 10.3      & 4.4 & 4.9  & $\sim$1.9 & $\sim$2.2 & 20.4-20.5  & 19.2       \\
(434326) 2004 JG$_6$ & 2021-05-29               & evening         & 43.3         & 40.7                & 9.4                & 62.3      & 0.8 & 1.1  & 2.0       & 2.1       & 20.0        & 19.7       \\
P/2021 N1 (ZTF)         & 2021-06-17               & morning         & 53.2         & -54.7               & 5.6                & 56.6      & 1.0 & 1.1 & 2.0-2.5 & 2.0-2.2 & 20.4-20.7  & 17.6-19.2   \\
(594913) \an & 2021-08-04 to 2022-02-14 & morning         & 39.9         & -41.6               & -0.5               & 87.3      & 0.7& 0.8  & 2.4-2.8 & 2.1-2.9 & 19.4-20.4  & 17.8-18.6  \\
2021 PB$_2$           & 2021-09-13               & morning         & 46.3         & -47.9               & 4.5                & 79.7      & 0.7 & 0.8  & 2.1       & 2.0       & 20.8        & 20.2       \\
C/2021 QM45 (PANSTARRS)       & 2022-02-14 to 2022-08-06 & evening         & 41.8         & 39.6                & 7.8                & 11.4      & 3.3 & 4.0  & 1.7-2.7 & 1.6-2.3 & 19.9-20.9  & 18.4-19.2  \\
\end{tabular}
\begin{tablenotes}
\item \textbf{Notes.} (1) Recovery date range, (2) time of twilight, (3) Sun-centric angular distance of first recovery observation, (4) Sun-centric ecliptic longitude of first recovery observation, (5) Sun-centric ecliptic latitude of first recovery observation, (6) phase angle of first recovery observation, (7) heliocentric distance of first recovery observation, (8) geocentric distance of first recovery observation, (9) airmass range of recoveries, (10) seeing range of recoveries, (11) limiting magnitude range of recovery observations in r band, (12) PSF-fitted r magnitude range of recoveries.
\end{tablenotes}
\end{table}
\end{landscape}

\end{document}